\documentclass[floats,floatfix,showpacs,amssymb,prd,twocolumn,superscriptaddress,nofootinbib,nolongbibliography,reprint]{revtex4-1}

\usepackage{amssymb,amsmath,verbatim,mathtools,needspace,enumitem,etoolbox,graphicx,physics,microtype,afterpage,xspace,tabularx,lmodern,multirow}
\usepackage{gensymb}
\usepackage[dvipsnames, usenames]{xcolor}
\definecolor{linkcolor}{rgb}{0.0,0.3,0.5}
\usepackage[unicode, colorlinks=true, linkcolor=linkcolor, citecolor=linkcolor, filecolor=linkcolor, urlcolor=linkcolor, linktocpage, breaklinks]{hyperref}
\usepackage[all]{hypcap}
\usepackage[T1]{fontenc}
\usepackage[utf8]{inputenc}
\usepackage[usenames,dvipsnames]{xcolor}
\usepackage{csquotes}
\hypersetup{colorlinks=true,citecolor=romared,linkcolor=romared,urlcolor=romared}

\definecolor{romared}{RGB}{142,0,28}

\newcommand{\be}{\begin{equation}}
\newcommand{\ee}{\end{equation}}

\def\be{\begin{equation}}
\def\ee{\end{equation}}
\newcommand{\beq}{\begin{eqnarray}}
\newcommand{\eeq}{\end{eqnarray}}
\usepackage{makecell}
\usepackage{soul}

\newcolumntype{Y}{>{\centering\arraybackslash}X}

\begin{document}

\title{Stability of the fundamental quasinormal mode in time-domain
  observations against small perturbations}

\author{Emanuele Berti}
\affiliation{William H. Miller III Department of Physics and Astronomy, Johns Hopkins University, 3400 North Charles Street, Baltimore, Maryland, 21218, USA}
\author{Vitor Cardoso} 
\affiliation{Niels Bohr International Academy, Niels Bohr Institute, Blegdamsvej 17, 2100 Copenhagen, Denmark}
\affiliation{CENTRA, Departamento de F\'{\i}sica, Instituto Superior T\'ecnico -- IST, Universidade de Lisboa -- UL, Avenida Rovisco Pais 1, 1049-001 Lisboa, Portugal}
\author{Mark Ho-Yeuk Cheung}
\affiliation{William H. Miller III Department of Physics and Astronomy, Johns Hopkins University, 3400 North Charles Street, Baltimore, Maryland, 21218, USA}
\author{Francesco~Di~Filippo} 
\affiliation{Center for Gravitational Physics and Quantum Information,
Yukawa Institute for Theoretical Physics, Kyoto University, Kyoto 606-8502, Japan}
\author{Francisco Duque} 
\affiliation{CENTRA, Departamento de F\'{\i}sica, Instituto Superior T\'ecnico -- IST, Universidade de Lisboa -- UL, Avenida Rovisco Pais 1, 1049-001 Lisboa, Portugal}
\author{Paul Martens} 
\affiliation{Center for Gravitational Physics and Quantum Information,
Yukawa Institute for Theoretical Physics, Kyoto University, Kyoto 606-8502, Japan}
\author{Shinji Mukohyama} 
\affiliation{Center for Gravitational Physics and Quantum Information,
Yukawa Institute for Theoretical Physics, Kyoto University, Kyoto 606-8502, Japan}
\affiliation{Kavli Institute for the Physics and Mathematics of the Universe (WPI),
The University of Tokyo, Chiba 277-8583, Japan}
\begin{abstract}
  Black hole spectroscopy with gravitational waves is an important tool to measure the mass and spin of astrophysical black holes and to test their Kerr nature. Next-generation ground- and space-based detectors will observe binary black hole mergers with large signal-to-noise ratios and perform spectroscopy routinely. It was recently shown that small perturbations due, e.g., to environmental effects (the \enquote{flea}) to the effective potential governing gravitational-wave generation and propagation in black hole exteriors (the \enquote{elephant}) can lead to arbitrarily large changes in the black hole's quasinormal spectrum, including the fundamental mode, which is expected to dominate the observed signal. This raises an important question: is the black hole spectroscopy program robust against perturbations? We clarify the physical behavior of time-domain signals under small perturbations in the potential, and we show that changes in the {\it amplitude} of the fundamental mode in the prompt ringdown signal are parametrically small. This implies that the fundamental quasinormal mode extracted from the observable time-domain signal is stable against small perturbations. The stability of overtones deserves further investigation.
\end{abstract}

\hfill YITP-22-46, IPMU22-0025

\maketitle

\section{Introduction}\label{sec:intro}

Recent progress in gravitational-wave astronomy~\cite{LIGO:2016,LIGO:2020} and very long baseline interferometry~\cite{EHT:2019} puts black holes (BHs) at the center of an intense observational program to constrain or detect signatures of new physics beyond general relativity (GR)~\cite{Berti:2015itd,Cardoso:2017cqb,Berti:2018cxi,Berti:2018vdi,Barack:2018yly,Carballo:2018,Cardoso:2019rvt}. In particular, BH spectroscopy is expected to play a pivotal role in this program~\cite{Detweiler:1980gk,Dreyer:2003bv,Berti:2005ys,LIGO:2016lio}. The late-time gravitational-wave signal produced by a binary merger (the so-called ``ringdown''~\cite{Kokkotas:1999bd,Berti:2009kk}) is described by a superposition of damped exponentials with complex frequencies, known as the quasinormal modes (QNMs) of the system, whose detection can be used to test the predictions of GR. This is only possible if the QNMs are spectrally stable, otherwise small environmental perturbations could produce large deviations in the QNM spectrum and hide any hypothetical signatures of new physics.  

Recent work analyzing spectral stability through calculations of the pseudospectrum~\cite{Jaramillo:2020tuu,Jaramillo:2021tmt,Destounis:2021lum,Gasperin:2021kfv} confirmed early predictions that the QNM spectrum should, in fact, be spectrally unstable under small perturbations~\cite{Nollert:1996rf}. In particular, the fundamental QNM frequency can have corrections of order one when a tiny perturbation is added to the potential~\cite{Cheung:2021bol}.
Physical perturbations of the potential can have various origins, including nonlinear effects within vacuum GR~\cite{Gleiser:1998rw,Campanelli:1998jv,Zlochower:2003yh,Nakano:2007cj,Ioka:2007ak,Okuzumi:2008ej,Pazos:2010xf,Sberna:2021eui}, ordinary matter~\cite{Leung:1997was,Leung:1999iq,Barausse:2014tra}, dark matter~\cite{Chung:2021roh}, or modifications of Einstein's theory of gravity~\cite{Cardoso:2019mqo,McManus:2019ulj}.
In this paper we address an important question: could this instability affect our ability to do BH spectroscopy with gravitational-wave observations? The key point here is that the instability refers to calculations of the spectrum in the frequency domain, which can be misleading. The time- and frequency-domain problems are simply related by a Fourier transformation, and hence are completely equivalent. In the frequency domain analysis, however, one usually focuses on the spectrum of the relevant operator, which only describes the late-time response. Therefore, to observe the equivalence it may be necessary to observe the system for a very long time and with very high precision. For small perturbations, the observation times and signal-to-noise ratios required to establish this equivalence are probably beyond current observational capabilities. This was already shown in the context of piecewise approximations to the potential~\cite{Nollert:1996rf, Daghigh:2020jyk}, environmental perturbations~\cite{Barausse:2014tra} and horizonless exotic compact objects (ECO)~\cite{Cardoso:2019rvt}. In ECOs, the BH horizon is substituted by a (partially or totally) reflecting surface. The spacetime geometry is not modified outside this surface, and it is usually assumed that the dynamics of the perturbations is unchanged with respect to the BH case. The QNM spectrum obtained by imposing reflective boundary conditions at the ECO surface is very different from the spectrum of a BH. However, by causality, the time-domain response of a BH and an ECO is exactly the same for the time necessary for the perturbation to propagate to the surface of the ECO and then be reflected back~\cite{Cardoso:2016rao} (see e.g.~\cite{Mark:2017dnq,Hui:2019aox}, where this is understood through an expansion of the Green's function in terms of echoes). Frequency-domain studies also pointed out that ECOs would have low-frequency oscillation modes, which could be excited by orbiting bodies~\cite{Cardoso:2019nis,Maggio:2021,Sago:2021iku}, but a time-domain analysis of these systems shows that the timescale to excite the resonances is typically much longer than the gravitational-wave evolution timescale~\cite{Cardoso:2022fbq}.

The goal of this paper is to investigate whether the instability of the fundamental QNM obtained in~\cite{Cheung:2021bol} translates into an instability of the mode as observed (either theoretically or in actual experiments) in a time-domain analysis.
In Sec.~\ref{sec:Setup} we introduce our parametrization of the perturbations of the potential.
In Sec.~\ref{sec:Numerical} we describe our numerical codes, and in Sec.~\ref{sec:results} we give the main results of our analysis.
In Sec.~\ref{sec:discussion} we present some concluding remarks and directions for future work.
Throughout the paper we set $G=c=1$.

\section{Perturbations of the potential}\label{sec:Setup}

We will study a modified version of the wavelike equations that govern gravitational perturbations around nonrotating BHs in GR~\cite{Regge:1957td, Zerilli:1970wzz}, which have the general form\footnote{More in general, we could include a source term that depends on what is causing the perturbation. We have also performed a study of the nonhomogeneous equation and found no qualitative differences. For clarity and simplicity of presentation, in this work we focus exclusively on the homogeneous equation.}
\beq
\frac{\partial^2 \Psi}{\partial r_*^2} - \frac{\partial^2 \Psi}{\partial t^2} - V \Psi = 0\,,
\label{eq:MasterEq}
\eeq
where $\Psi$ is a complex ``master function,''  the tortoise coordinate $r_*$ is defined in terms of the usual Schwarzschild areal coordinate $r$ by $dr/dr_*=1-1/r$,
and $V=V(r)$ is a radial potential given by 
\beq
V = V_\text{0} +  \epsilon V_\text{bump} \, ,
\label{eq:Potential}
\eeq
Here and below we measure lengths in units of the Schwarzschild radius (i.e., we set $2M=1$).
We denote by $V_\text{0}$ the unperturbed potential for the odd parity (or Regge-Wheeler~\cite{Regge:1957td}, $V_\text{0}=V_\text{RW}$) and even parity (or Zerilli~\cite{Zerilli:1970se,Zerilli:1970wzz}, $V_\text{0}=V_\text{Z}$) perturbations, while the second term $\epsilon V_\text{bump}$ represents a small ($\epsilon \ll 1$) perturbation or ``bump.'' Equations~\eqref{eq:MasterEq}-\eqref{eq:Potential} arise naturally when describing
families of different background spacetimes parametrized by some small quantity $\epsilon$. In this context, $\epsilon$ could stand, for example, for the BH charge, in which case the procedure to reduce the relevant perturbation equations to the form above is outlined in Ref.~\cite{Cardoso:2019mqo}. 
Alternatively, an $\mathcal{O}(\epsilon)$ modification of general relativity may also result in an $\mathcal{O}(\epsilon)$ perturbation of the potential, provided that the speed limit of gravitational waves is unchanged.

\begin{figure}[h]
    \centering
    \includegraphics[width=\columnwidth]{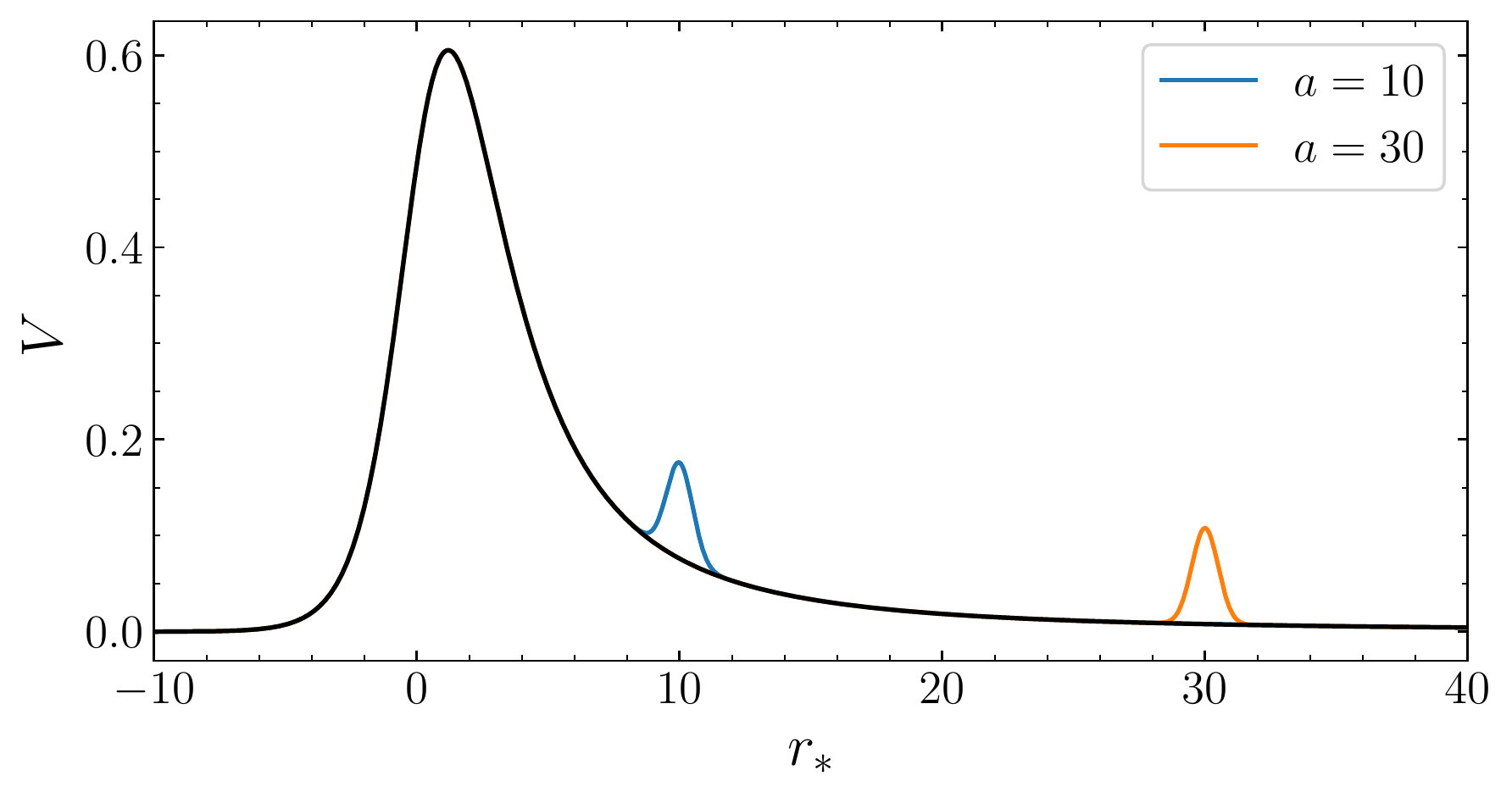}
    \caption{The unperturbed ($\epsilon=0$) and perturbed potentials used in this study. The Regge-Wheeler potential $V_\text{RW}$ with $l = 2$ is shown in black, while two perturbative bumps $\epsilon V_\text{bump}$ with $\epsilon = 0.1$ are shown in blue (for the case of $a = 10$) and orange ($a=30$). The unperturbed potential has a peak close to the light ring, at $r\simeq 1.5$.
  }%
    \label{fig:Potential}
\end{figure}

We will mostly focus on the odd-parity case with angular number $l = 2$, and thus the unperturbed potential is~\cite{Regge:1957td}
\beq
V_\text{RW} = 3 \left(1-\frac{1}{r}\right)\left[\frac{2}{r^2} - \frac{1}{r^3} \right]  \, .
\label{eq:RW}
\eeq
Following Ref.~\cite{Cheung:2021bol}, we will assume that the bump is localized around some radius $r_* = a$ and that it goes to zero faster than $V_\text{RW}$ as $r_* \rightarrow \infty$. We will study for concreteness a Gaussian bump of the form
\beq
V_\text{bump} = \exp\left[-\frac{(r_*-a)^2}{2\sigma^2}\right] \, .\label{bump_width}
\eeq
Some example potentials are shown in Fig.~\ref{fig:Potential}.
The specific form of the bump in Eq.~\eqref{bump_width} is chosen as a simple illustrative example. However, we have considered different types of perturbation, such as (i) a bump which is concave rather than convex, corresponding to negative $\epsilon$, and (ii) a bump which decays as $r^{-3}$ at large distances and which is exactly zero for small $r$, mimicking a perturbation due to a thin shell of matter. 
The qualitative behavior of the time-domain signal is the same in all these different cases, so we will not report the results here. In fact, the only relevant aspect of the perturbation is that it introduces a second small peak. When this is not the case (e.g., when $a$ is very small) the analysis of Refs.~\cite{Cheung:2021bol,Cardoso:2021wlq} confirms that the instability is not present.  Note that the isospectrality of the Zerilli and Regge-Wheeler potentials is broken even if we use the {\em same} perturbative bump in both equations~\cite{Cardoso:2019mqo,McManus:2019ulj} (although in realistic astrophysical scenarios the odd- and even-parity perturbative ``bumps'' are not expected to be the same).

\section{Numerical framework}\label{sec:Numerical}

\subsection{Time domain}\label{subsec:td}

Let us start by numerically solving the axial-type version of Eq.~\eqref{eq:MasterEq} in the time domain.  We prescribe initial data corresponding to a localized Gaussian pulse
\begin{subequations}
  \label{id_gaussian}
  \begin{align}
    &\Psi(t=0,r) = 0 \, , \\
    &\frac{\partial \Psi}{\partial t}(t=0,r) = \exp\left[-\frac{(r_*-5)^2}{2}\right] \, . 
  \end{align}
\end{subequations}
where again all dimensionful quantities are given in units of the BH's Schwarzschild radius ($2M=1$).

We have performed the numerical integration with several independent codes. The first code employs a hyperboloidal compactification~\cite{PanossoMacedo:2019npm, Zenginoglu:2011zz}. The gauge freedom in GR allows us to switch from the original $(t,r_*)$ coordinates to a set of hyperboloidal compactified coordinates $(\tau, \rho)$ defined by the relations
\beq
r_* &=& \frac{\rho}{\Omega(\rho)} \, ,\qquad \tau =t -\frac{\rho}{\Omega(\rho)} - \rho \, , \\
\Omega &=& 1 - \left(\frac{\rho - R_* }{S - R_*} \right)^4 \Theta \left(\rho - R_* \right) \,.
\eeq
This transformation maps the infinite domain of the radial tortoise coordinate to a compactified domain, where the outer boundary is located at $\rho=S$ and corresponds to the null infinity $\mathcal{I}^+$. Here $R_*$ is a transition radius between an interior domain, where $\rho = r_*$, and the hyperboloidal layer that approaches $\mathcal{I}^+$. 
After prescribing initial data for $\Psi$, Eq.~\eqref{eq:MasterEq} is evolved in time using a two-step Lax-Wendroff algorithm with second-order finite differences. For a complete description of this numerical code, we refer the reader to Refs.~\cite{Krivan:1997hc,Pazos-Avalos:2004uyd,Zenginoglu:2011zz,Cardoso:2021vjq,Cardoso:2021wlq}. 
The other codes that we have implemented to cross-check results directly solve Eq.~\eqref{eq:MasterEq} on a uniformly spaced $r_*$-domain without any change of coordinates. The dynamical variables are integrated in time $t$ using either the fourth-order Runge--Kutta method or the iterated Crank-Nicolson method with two iterations. One of these codes was previously used in Refs.~\cite{Bhattacharjee:2018nus,Mukohyama:2020lsu}. 

\begin{figure*}[t]
    \centering
    \includegraphics[width=\textwidth]{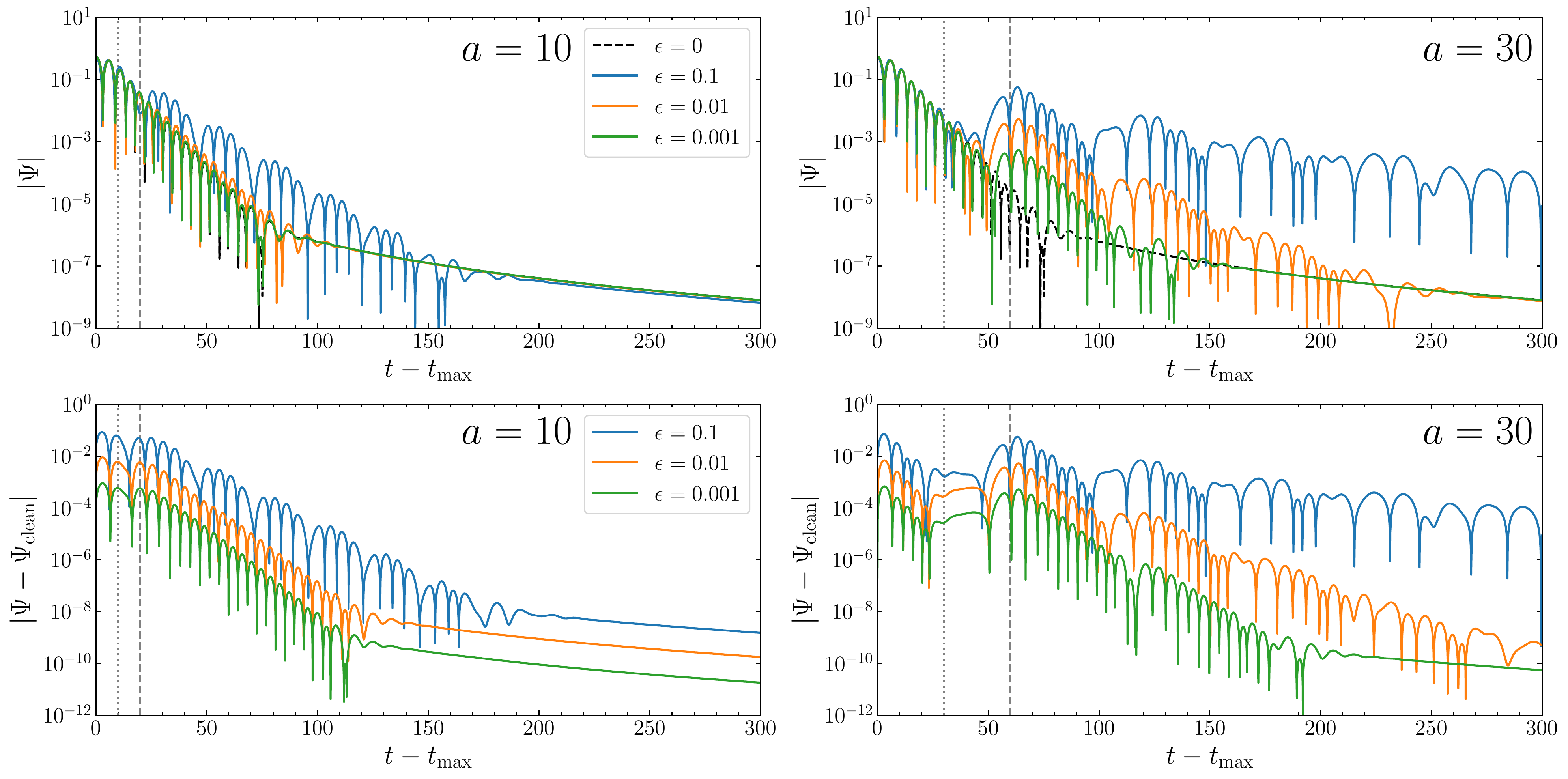}
    \caption{Top panels: absolute value of the time-domain waveform arising from the scattering of the Gaussian pulse of Eq.~\eqref{id_gaussian} for ``bumps'' with different amplitudes $\epsilon$, located at two selected distances $a$ from the main peak.
      The bump width $\sigma$ in Eq.~\eqref{bump_width} is fixed at $\sigma = 0.5$. 
    ``Echoes'' are apparent when the bumps are located at large distances ($a = 30$).
    The dotted and dashed vertical gray lines correspond to $t - t_\text{max} = a$ and $2a$, and they illustrate how the delay between echoes is related to the size of the ``cavity'' between the two maxima in the perturbed potential.
    Bottom panels: absolute value of the difference between the waveforms shown in the top panels and the unperturbed clean waveform without a bump ($\epsilon = 0$).} 
    \label{fig:Psis_epsvar}
\end{figure*}

At late times, the time-domain signal decays as a linear combination of exponentially damped sinusoids, whose frequencies and damping times can be extracted by fitting the waveform with the $N$-mode template
\beq
\Psi(t) &=& {\rm Re} \sum_{n=0}^{N-1} A_n e^{-i (\omega_n t - \phi_n)} \label{eq:modes}\\
&=& \sum_{n=0}^{N-1} A_n e^{\omega_{n I} t} \cos(\omega_{n R} t - \phi_n)\, ,
\label{eq:template}
\eeq
where the index $n$ labels the different modes we find by fitting and it does not necessarily coincide with the overtone number. Each mode is characterized by four parameters: an amplitude $A_n$, a phase $\phi_n$, and the real and imaginary parts of the QNM frequency $\omega_n = \omega_{nR} + i\omega_{nI}$.
In Sec.~\ref{sec:results} we will find that several QNMs could have similar decay times, and hence comparable amplitudes. In this situation, a good fit of the waveform requires a relatively large number of modes $N$. The largest number of modes we will look for is $N = 8$, corresponding to $8 \times 4 = 32$ fitting parameters. We found that the \texttt{NonLinearModelFit} function provided by \textit{Mathematica} works well for this high-dimensional nonlinear fitting problem. We have verified the quality of the fits by using toy models and by cross-checking with other fitting methods, and we checked that the fit residues are small when the fitted frequencies converge well.

\subsection{Frequency domain: The characteristic QNMs}\label{subsec:fd}

The eigenfrequencies $\omega_n$ of Eq.~\eqref{eq:modes} can be computed directly from Eq.~\eqref{eq:MasterEq} with a Laplace transform, i.e., by solving the equation
\begin{equation}\label{eq:MasterFD}
\frac{\partial^2 \Psi}{\partial r_*^2} +\left(\omega^2 - V\right)\Psi =0\,.
\end{equation}
The QNM frequencies $\omega_n$ correspond to the poles of the Green's function of Eq.~\eqref{eq:MasterFD} with the appropriate boundary conditions of ingoing waves at $r_* = -\infty$, and outgoing waves at $r_* = +\infty$.
The frequencies can be found by a shooting method~\cite{Chandrasekhar:1975zza}: we starting from each of the two boundaries and numerically integrate inwards or outwards iteratively searching for the roots of the Wronskian, i.e., the values of $\omega$ for which the two solutions match smoothly in an intermediate region.

We have performed such a direct-integration analysis using a modification of the \textit{Mathematica} notebook used in Ref.~\cite{Molina:2010fb} and available online~\cite{RDwebsites}.

\section{Results}\label{sec:results}

The fundamental mode of the Regge-Wheeler equation is unstable if the perturbative bump is sufficiently large or if it is located sufficiently far away~\cite{Cheung:2021bol}.
By definition, we will say that ``destabilization'' occurs when a quantity that characterizes the BH's response (e.g., the QNM frequency or the waveform amplitude) changes by an amount much larger than the magnitude of the perturbation.
Reference ~\cite{Cheung:2021bol} investigated the spectral instability of the fundamental mode. Here we will show that while the fundamental mode is indeed spectrally unstable, the change in the time-domain waveform amplitude is of the same order as the size of the perturbation. In this sense, the time-domain waveform itself is stable under perturbations. Furthermore, the {\em observed} QNM frequency is also stable in the sense defined above.
This is because, as we will see, the {\em early part} of the waveform is not significantly affected by the perturbation. 

\begin{figure}[h]
    \centering
    \includegraphics[width=\columnwidth]{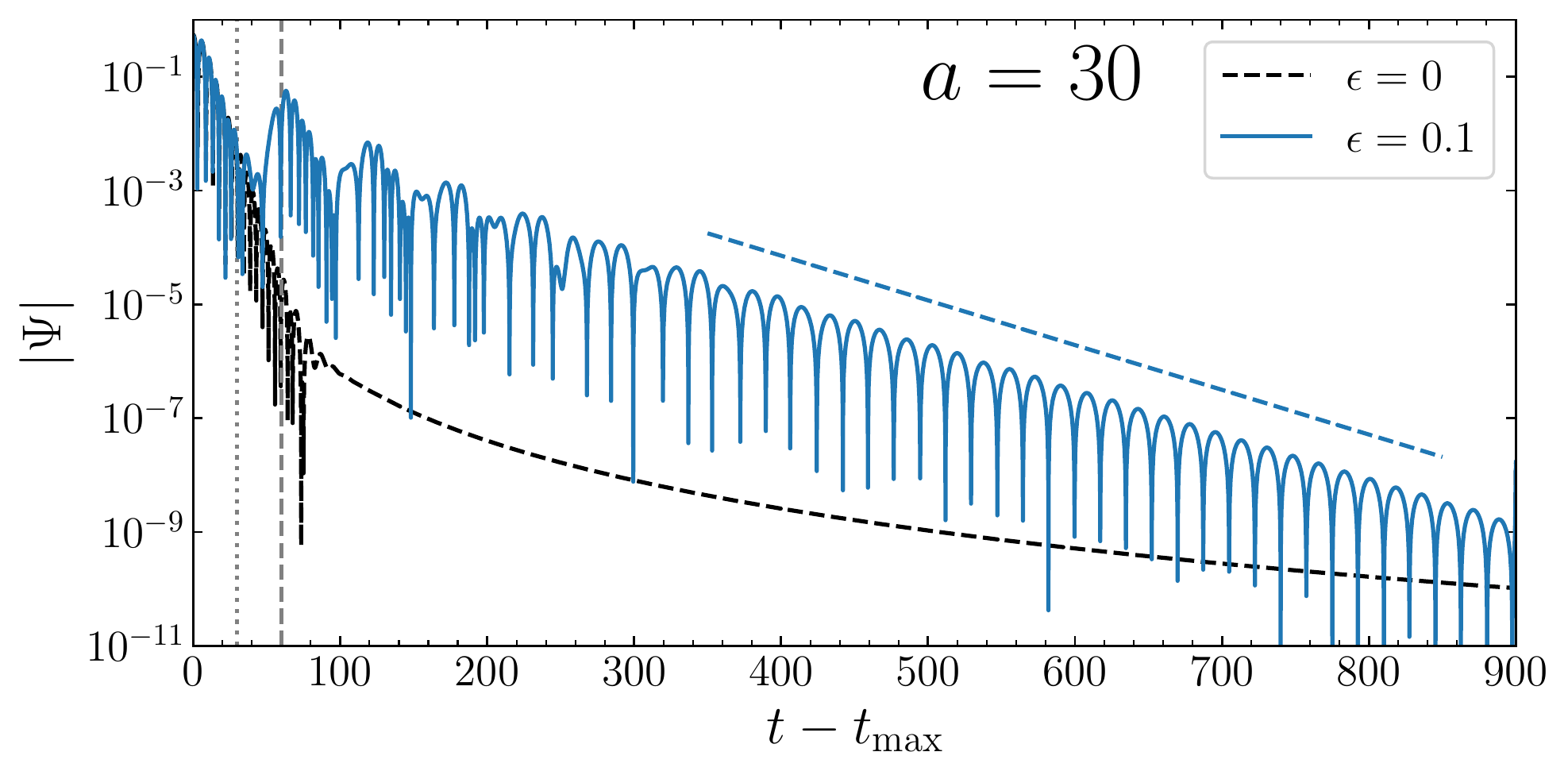}
    \caption{Time-domain waveform for $a = 30$ and $\epsilon = 0.1$ over a longer time span (up to $t - t_{\rm max} = 900$):
    the waveform transitions to a damped sinusoid corresponding to the new fundamental mode at late times.
    For comparison, the blue dashed line shows the expected decay time of the new fundamental mode, which corresponds to the bottom blue cross with smallest $|\omega_{nI}|$ in the top panel of Fig.~\ref{fig:RW_a_30_modes} below.
    }%
    \label{fig:Psis_epsvar2}
\end{figure}

\subsection{Stability of the prompt ringdown signal and the destabilization mechanism}
In Fig.~\ref{fig:Psis_epsvar} we show the waveform generated by the scattering of Gaussian pulses. We define $t_{\rm max}$ to be the time at which $|\Psi|$ is at its maximum value, and we only show the waveform from $t_{\rm max}$ on-wards. For the ``clean'' case in the absence of a perturbing bump (i.e., for $\epsilon=0$) we see a familiar exponentially decaying ringdown waveform followed by a power-law tail. Bumps with small values of $\epsilon$ cause correspondingly small modifications in the waveform at early times. As seen in the lower panels of Fig.~\ref{fig:Psis_epsvar}, at times before the first echo ($t - t_{\rm max} \lesssim a$), the absolute value of these modifications scales with $\epsilon$: in other words, bumps in the potential of amplitude $\epsilon$ induce changes in the gravitational-wave signal that are also of order $\epsilon$. This is one of our main results.
The prompt BH ringdown signal close to the peak of the waveform observed by gravitational-wave detectors is affected by a small environmental disturbance, but the modification in the prompt ringdown is not expected to be observable at the signal-to-noise ratios achievable by current detectors, consistently with previous claims~\cite{Barausse:2014tra,Cardoso:2019rvt}.

Figure~\ref{fig:Psis_epsvar} also illustrates how the fundamental mode destabilization is realized in the time domain~\cite{Cheung:2021bol}. When a perturbatively small bump is added to the potential, the {\it frequency content} of the waveform at late times is, indeed, drastically different from the $\epsilon=0$ case. The late-time waveform is well-described by a superposition of QNMs which are very different from (and longer-lived than) the fundamental mode of the BH spacetime in the absence of perturbations. This is most evident in the waveform with $\epsilon = 0.1$, which is shown over a longer time interval in Fig.~\ref{fig:Psis_epsvar2}: at late times the waveform decays with a QNM frequency which has lower frequency and longer damping time than the fundamental QNM of the Schwarzschild spacetime. For smaller values of $\epsilon$ the waveform amplitude is modified by a smaller amount, and this difference in the late-time behavior is partially masked by the familiar Price power-law tail observable in the unperturbed (black dashed) waveform.

In general, when we add a large bump in the potential the QNMs have a longer damping time, and hence they survive longer before ``diving below the tail.'' 
When the perturbative bump is located far away from the original potential peak (see e.g. the green and orange lines for $a=30$ in the right panels of Fig.~\ref{fig:Psis_epsvar}), we first observe lower amplitude copies (echoes) of the original pulse, which eventually give way to a different ringdown signal, characterized by frequencies and damping times which are different from the $\epsilon=0$ case. This behavior is easily interpreted. For large $a$, there is a clear separation of timescales between the ringdown pulse produced at the photonsphere (the peak of the $\epsilon=0$ potential) and the light travel time characterizing the ``cavity'' located between the photonsphere and the bump. Thus, what we have is a pulse bouncing back and forth within the cavity and gradually losing its high-frequency component, which tunnels out more easily. This produces a sequence of echoes repeating at a characteristic frequency defined by the cavity size and damped on a timescale defined by the transmission coefficient of the small peak, as shown in Fig.~\ref{fig:Psis_epsvar} (see also Refs.~\cite{Cardoso:2016rao,Cardoso:2016oxy,Cardoso:2017cqb,Cardoso:2019rvt} for a very similar behavior when the bump is arbitrarily close to the horizon). These two scales determine the QNM spectrum of the bumpy potential, which can be nonperturbatively different from the $\epsilon=0$ case. A simple rule of thumb for the echoes to be visible is that the prompt ringdown lifetime $\sim 9\sqrt{3}M$ (allowing for three e-folding times) should be smaller than the travel time within the cavity $\sim 2a$~\cite{Cardoso:2017cqb,Cardoso:2019rvt}, and therefore we should require $a\gtrsim 4$ (in units $2M=1$). Note that, however, the {\it amplitude} of the echoes and of the induced QNM ringing is proportional to $\epsilon$.

Let us stress once more how destabilization affects the signal. The QNM frequencies can be destabilized, i.e. their relative variation can be much larger than $\epsilon$ when a perturbative bump is included in the potential. However the change in the amplitude of the waveform is still proportional to $\epsilon$ for the early ringdown, as shown in the bottom panels of Fig.~\ref{fig:Psis_epsvar}. In this sense, the waveform changes by an amount proportional to the size of the bump: although the QNM frequencies are destabilized, an arbitrarily small bump will not alter the prompt ringdown waveform by a significant amount. 

Our results can be interpreted in terms of the shape of the perturbed potential shown in Fig.~\ref{fig:Potential}.  If $V_\text{bump}$ is localized close to the peak of $V_0=V_\text{RW}$ the QNM frequencies should change, because the bump effectively changes the shape of the original potential.
If instead $V_\text{bump}$ is localized far away from the peak of $V_0$, the QNM frequencies should also be different, because they depend on the behavior of the potential in the entire radial domain. Nevertheless, in this regime we would expect that the early-time waveform should be characterized by the QNM frequencies of the unperturbed potential, because the bump is too far away to modify the unperturbed BH spectrum (see also the discussion in Ref.~\cite{Cardoso:2019rvt} for bumps located ``behind'' the peak). In this regime one can think of the prompt ringdown as being excited at the light ring and propagating outwards, but this wave train must tunnel out of the bump to reach asymptotic observers. The reflection coefficient for tunneling depends on the frequency, but is  $\mathcal{O}(\epsilon)$ for all frequencies, and therefore it can modify the prompt ringdown wave train (excited at the light ring) only by a small factor $\mathcal{O}(\epsilon)$.

While the relative modification $|\Psi - \Psi_{\rm clean}| / |\Psi|$ of the early ringdown scales as $\epsilon$, we observe larger changes in the waveform during the echo-dominated phase. Eventually the power-law tail dominates at late times, and the relative modification scales as $\epsilon$ again.
The larger order-of-magnitude modification can be attributed to the echoing ringdown waves.
The amplitude of the $n$th echo is larger than the prompt ringdown (the ``$0$th'' echo) signal roughly by a factor $\tilde{\epsilon}^n$, where $\tilde{\epsilon} \sim \epsilon e^{2a|\omega_{0I}|}$.
This is because a bounce in the cavity between the two bumps occurs on a timescale of order $2a$, during which the wave in the cavity does not decay but the $0$th signal keeps decaying as $e^{-|\omega_I|t}$, and each reflection at the $\epsilon$-bump reduces the amplitude of the wave in the cavity by a factor $\sim \epsilon$.
Then, the modification of the waveform during the echo phase at a given $t$ ($>2a)$ is not of order $\epsilon$, but rather of order $\tilde{\epsilon}^n$, with $n\sim t/(2a)$.
At even later times the power-law tail dominates the waveform, and the modification to the waveform is again of order $\epsilon$. This can be understood as follows. One can imagine placing a very wide bump from which it is very hard to tunnel out, so the fluctuation would be effectively confined. Then the relative modification in the waveform (with respect to vacuum) would diverge, since vacuum fields obey radiative boundary conditions.
The same type of argument indicates that sufficiently low-frequency signals would be affected in a similar way, since they ``tunnel out'' poorly, and tunneling is exponentially sensitive to the barrier's height.
Now, the power-law tail is formally a zero-frequency signal, so its modification does not scale as $\tilde{\epsilon}^n$ (like the tunneling echo waves) but rather as $\epsilon$.

\subsection{Comparison between QNM frequencies and the time-domain signal}

We now wish to compare the QNM frequencies obtained by frequency-domain calculations (see Sec.~\ref{subsec:fd}) with those found by fitting the time-domain waveform with damped sinusoids using \textit{Mathematica}'s \texttt{NonLinearModelFit} function, as described in Sec.~\ref{subsec:td}. In this subsection we will compare the theoretical predictions in the time and in the frequency domain, while in the next subsection we will comment on the observational implications of these findings. 

When fitting the time-domain waveform, we are interested in two different regimes:

\begin{itemize}
\item[1)] We would like to understand whether the spectral instability affects the prompt ringdown radiation emitted by astrophysical BHs, which is typically louder and more easily detectable by gravitational-wave interferometers.
\item[2)] Even if the prompt ringdown is not affected, it is still theoretically interesting to understand whether the late-time waveform is characterized by the trapped-mode spectrum, i.e. by modes trapped in the cavity between the two peaks of the potential (see Fig.~\ref{fig:Potential}). 
\end{itemize}

\begin{figure}
    \centering
    \includegraphics[width=\columnwidth]{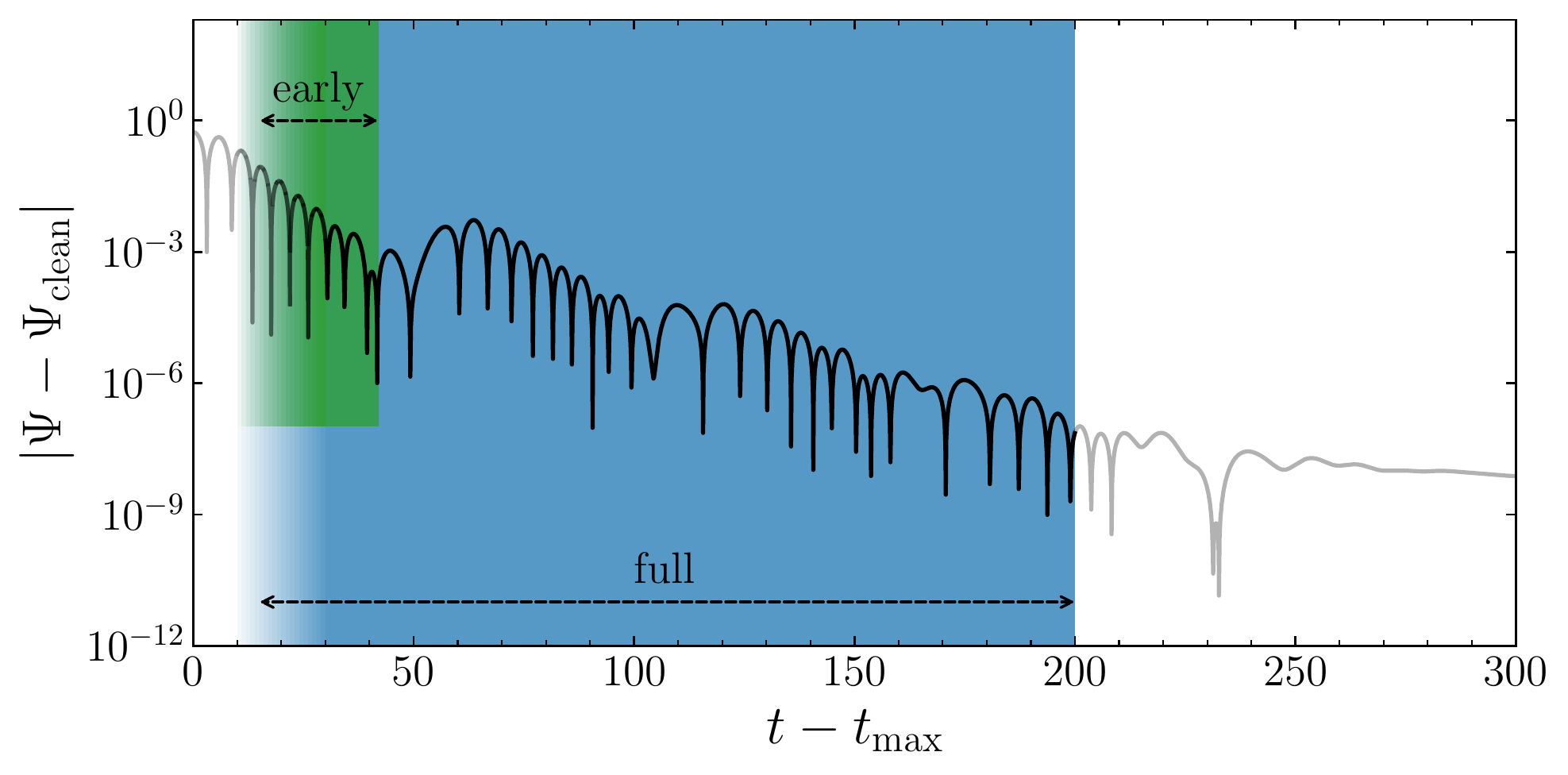}
    \caption{The section of the waveform we used for damped-sinusoid fitting.
    The waveform shown here refers to $a = 30$, $\epsilon = 0.01$, but the same fitting procedure applies in general to the other waveforms (with fitting ranges adjusted accordingly).}
    \label{fig:Psis_fit_range}
\end{figure}

\begin{figure}
    \centering
    \includegraphics[width=\columnwidth]{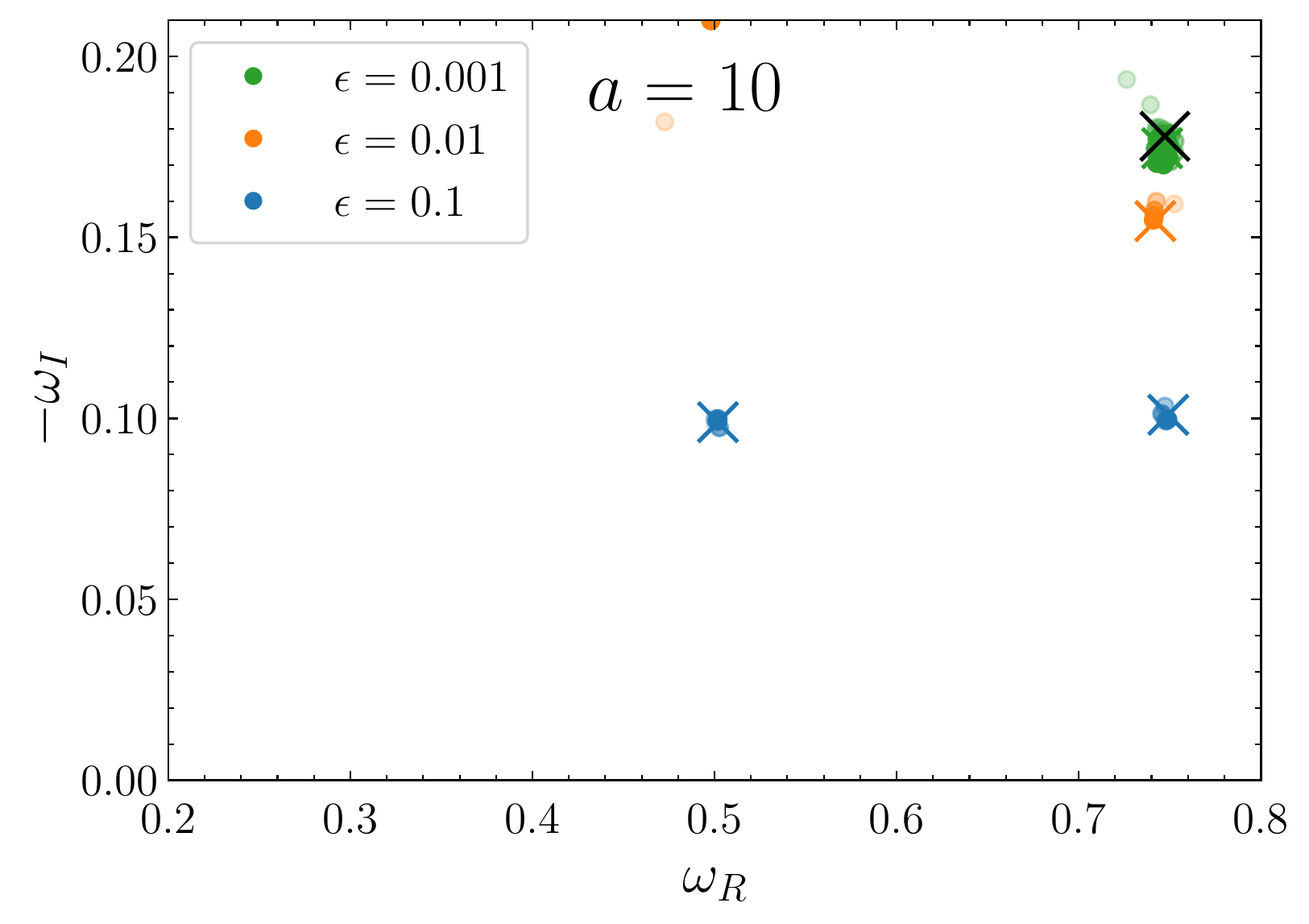}
    \caption{Comparison between the QNM frequencies obtained directly with the shooting method (crosses) and those by fitting the full time-domain waveform (dots). The black cross refer to the unperturbed clean fundamental mode.
    Different dots with the same color corresponds to fits with different starting times: the darker the dot, the later the starting time.
    The starting times are $t - t_{\rm max} = 10, 15, 20, 25, 30 M$ (cf. Fig.~\ref{fig:Psis_fit_range}).
    The orange dots visible at the top are due to a high-damping mode present above the plotting range. 
    }
    \label{fig:RW_a10_modes}
\end{figure}

We will start by investigating the late-time waveform. In Fig.~\ref{fig:Psis_fit_range} we highlight the two different portions of the waveform used to recover the QNM frequencies by our fitting procedure.
First of all, we discard times such that $t-t_{\rm max}\lesssim 5$ as they are contaminated by the direct propagation of the initial wave packet.

\begin{figure}
    \centering
    \includegraphics[width=\columnwidth]{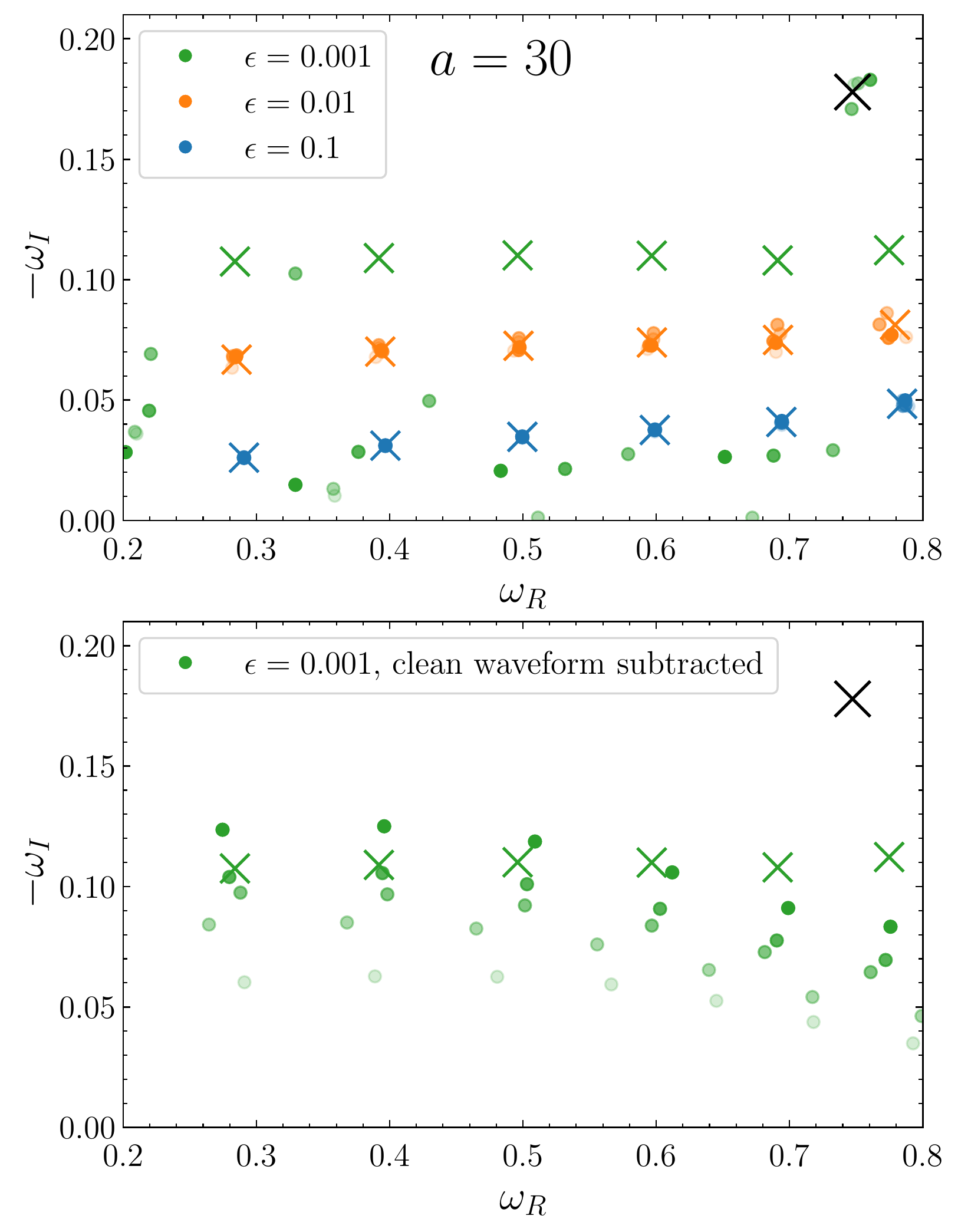}
    \caption{Top panel: same as Fig.~\ref{fig:RW_a10_modes}, but with $a = 30$.
    For $\epsilon = 0.001$, the full time-domain fits could only confidently detect a mode close to the fundamental mode of the unperturbed potential. This is because the ringdown wave amplitude becomes smaller than the late-time tail before we can clearly see the trapped modes.
    Bottom panel: we reconsider the troublesome case where $\epsilon = 0.001$, but we now subtract the unperturbed clean waveform from the actual waveform before fitting (i.e., we use the green curve in the bottom right panel of Fig.~\ref{fig:Psis_epsvar} for the fit).
    The QNM frequencies obtained by time-domain fitting do not converge precisely, but the structure of the QNM mode spectrum extracted from the residual waveform is in good agreement with the expected trapped QNM spectrum, especially for the real part. For both panels, as before, the starting times are $t - t_{\rm max} = 10, 15, 20, 25, 30 M$.
    }
    \label{fig:RW_a_30_modes}
\end{figure}

The trapped modes due to the potential bump shown (e.g.) in Figs.~\ref{fig:RW_a10_modes} and \ref{fig:RW_a_30_modes} where found by using the ``full'' fitting range (shaded in blue in Fig.~\ref{fig:Psis_fit_range}), i.e., we only discard the portion of the waveform which is significantly contaminated by the power-law tail.
To understand whether the destabilized QNM spectrum affects the prompt ringdown (as in Fig.~\ref{fig:RW_a30_50_modes} below), we only fit the ``early'' part of the waveform (shaded in green in Fig.~\ref{fig:Psis_fit_range}), where echoes do not affect the signal.
In each of these two fitting regimes (that is, within either the green or the blue shaded regions) we vary the starting time of the fit to make sure that the QNM spectrum recovered is insensitive to small changes in the fitting range.
The frequencies computed using different starting times are shown as dots of different shades in Figs.~\ref{fig:RW_a10_modes}--\ref{fig:RW_a30_50_modes}: the darker the dot, the later the starting time.
In all plots, crosses refer to the QNM frequencies found by the shooting method in the frequency domain. In particular, the black cross refers to the fundamental QNM of the unperturbed potential ($\epsilon=0$), while the unperturbed overtone QNMs are above the plotting range.

Consider first a bump located relatively close to the original potential peak ($a = 10$, Fig.~\ref{fig:RW_a10_modes}). Different colors refer to QNM frequencies obtained for different values of $\epsilon$ from the time-domain (dots) and frequency-domain calculations (crosses).
The frequencies extracted from time-domain fits cannot be expected to be arbitrarily accurate, because the waveform is contaminated by the direct propagation of the initial data, the power-law tail, and numerical noise, but they are still in very good agreement with the frequency-domain calculation for all the values of $\epsilon$ that we considered.

\begin{figure*}
    \centering
    \includegraphics[width=0.95\textwidth]{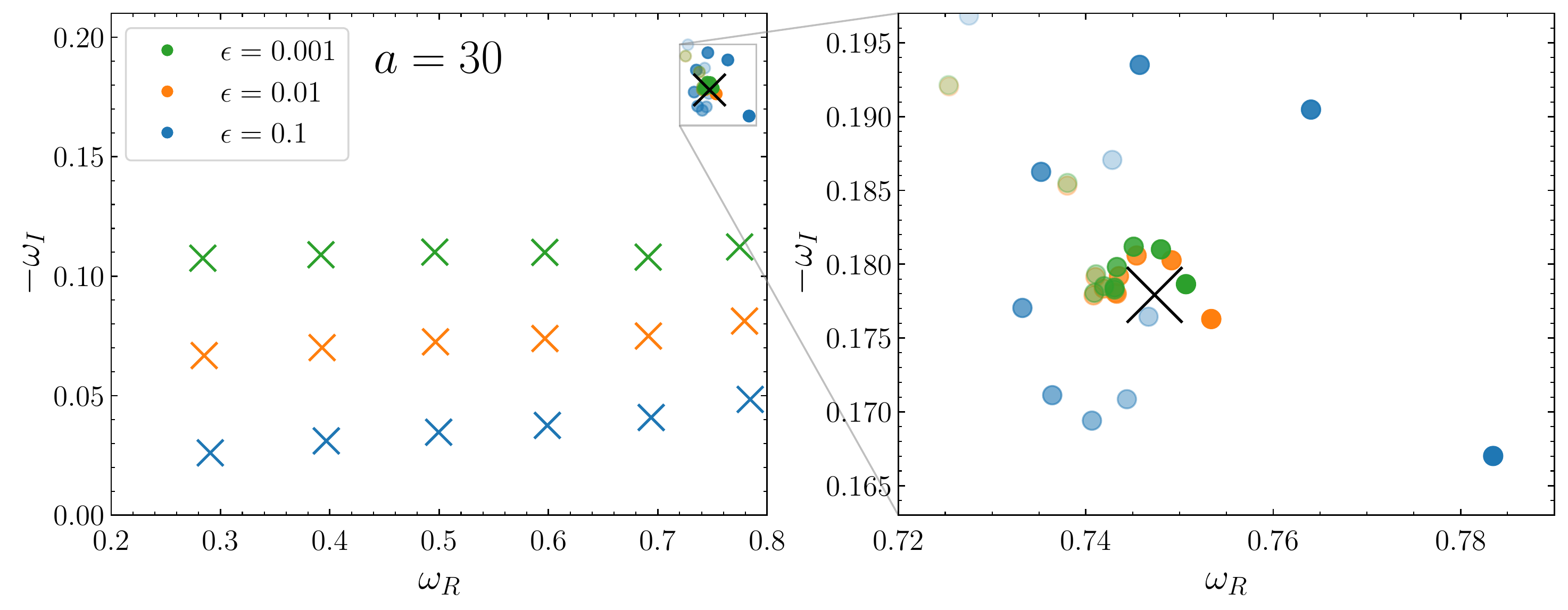}
    \caption{Same as Fig.~\ref{fig:RW_a10_modes}, but for $a = 30$, and we only fit the first train of the initial ringdown without echoes.  The starting times are $t - t_{\rm max} = 10, 11, 12,\dots, 20,  M$. All the dots obtained using the fitting method now cluster around the clean fundamental mode of the unperturbed potential. The zoom-in in the right panel shows that a perturbation of order $\epsilon$ can induce systematic errors (approximately of order $\epsilon$) in the measurement of the fundamental mode's frequency and damping time.}
    \label{fig:RW_a30_50_modes}
\end{figure*}

If the bump is farther away (say at $a = 30$, Fig.~\ref{fig:RW_a_30_modes}) the destabilization of the QNM frequencies is more evident.
As shown in Fig.~\ref{fig:Psis_epsvar}, at early times the waveform consists of a clean ringdown signal, but after one clear echo it transitions to a combination of ``new'' QNMs associated with the cavity, until their amplitude becomes so small that they are masked by the late-time power-law tail.
We include this intermediate transition regime in our fit by using the ``full'' fitting range as labelled in Fig.~\ref{fig:Psis_fit_range}.
The $\epsilon = 0.1$ is somewhat special because the decay time of the trapped QNMs is longer, so the waveform displays a clear transition to a new exponential decay dominated by a single trapped QNM before ``sinking'' below the power-law tail at very late times (see Fig.~\ref{fig:Psis_epsvar2}).
Given the large number of fitting parameters, it is quite remarkable that we can recover all of the slowest decaying QNMs as long as $\epsilon \gtrsim 0.01$, as shown in the top panel of Fig.~\ref{fig:RW_a_30_modes}.

The time-domain fit is more difficult when $\epsilon = 0.001$. In fact, the fits can only confidently detect a mode close to the clean fundamental mode of the unperturbed potential.
This is because the QNMs have a very short decay time for such small $\epsilon$, and the waveform does not have time to transition to the ``new'' trapped QNM spectrum before decaying below the power-law tail. Therefore the fitting algorithm can only pick up the clean mode, which is excited promptly at the light ring and is clearly observable at early times.
To remove the contribution of the tail, in the bottom panel of Fig.~\ref{fig:RW_a_30_modes} we subtract the clean ($\epsilon=0$) waveform from the signal and we repeat the fit using the green curve in the bottom-right panel of Fig.~\ref{fig:Psis_epsvar}, which as we can see contains more QNM oscillation periods that were previously hidden below the late-time power-law tail.
The fitted modes do not converge as in the cases with $\epsilon \geq 0.01$, but their general structure is now in good agreement with the trapped QNM spectrum computed in the frequency domain.
\subsection{Comparison between QNM frequencies and the prompt time-domain ringdown}
In realistic astrophysical scenarios, we would expect $\epsilon$ to be small. For example, the impact of matter effects on the potential for axial perturbations~\cite{Kokkotas:1999bd,Cardoso:2021wlq,Cheung:2021bol} is equivalent to adding a bump of amplitude $\epsilon V_{\rm bump}\sim \rho$, where $\rho$ is the matter density. If we express density in units of the BH mass, i.e.
\be
\rho M^2=1.6\times 10^{-18}\frac{\rho}{\rho_{\rm water}}\frac{M^2}{M_\odot^2}\,,
\ee
it is clear that $\epsilon$ should be small in most astrophysical scenarios. 
Moreover, as the signal-to-noise ratio of the ringdown in current detectors is relatively small~\cite{Berti:2016lat,Isi:2021iql,Ota:2021ypb,Bhagwat:2021kwv} and the signal decays exponentially, only the prompt ringdown would be expected to be observationally relevant.

For these reasons, in Fig.~\ref{fig:RW_a30_50_modes} we show again the QNM frequencies computed in the frequency domain and those recovered by fitting the waveform. This time however we restrict the fit to the prompt ringdown only, cutting the waveform at the beginning of the first echo. This portion of the waveform is well fitted by a {\em single} damped sinusoid, with the value of the frequency more or less converging to the location of the clean fundamental mode rather than the actual QNM frequencies present in the full signal. This is reassuring: the prompt ringdown signal is not appreciably affected by the perturbative bump. In other words, we should not expect the trapped QNMs to be observable in most astrophysically realistic scenarios.

\section{Discussion}\label{sec:discussion}

The main message of this paper is that spectral instability results obtained in the frequency domain should not be used naively in the interpretation of time-domain signals. 
In particular, the spectral instability of the fundamental QNM does not imply an instability of the ``physical'' fundamental QNM that dominates the prompt ringdown in the time-domain waveform.

For small values of $\epsilon$ and small separations of the perturbative bump $a$, prompt ringdown corrections relative to the fundamental QNM of the unperturbed potential are ${\cal O}(\epsilon)$. For larger values of $a$ the fundamental QNM is destabilized, as predicted by the frequency-domain analysis. However, if we restrict attention to the early phase of the ringdown (the only part that would be measurable in gravitational-wave observations at currently achievable signal-to-noise ratios), we would detect only the fundamental QNM of the unperturbed potential with corrections of ${\cal O}(\epsilon)$. At later times, the waveform displays echoes whose amplitude is suppressed by the $\mathcal{O}(\epsilon)$ reflection coefficient of the bump.  The first ringdown wave train is well described by the fundamental mode of the clean potential (modulo small corrections), whereas the ``true'' QNM spectrum only characterizes the system at late time. For small enough values of $\epsilon$, the amplitude is so suppressed that the ``true'' fundamental QNM  can be hidden by the polynomial late-time tail, or even by nonlinear corrections to the Regge-Wheeler (or Zerilli) equation. The frequency-domain instability is more effective for large values of $a$~\cite{Cheung:2021bol}. In the time domain, this means that we need to wait longer before the signal can be approximated by a damped sinusoid. In fact, for larger values of $a$ the number of distinguishable echoes increases.

In a mathematical sense, the frequency- and time-domain analyses are equivalent only if we have access to the full signal (i.e., to all values of the frequency, or to the waveforms at all times). This is practically unachievable in experiments. Our work shows that access to the late-time waveform is crucial to recover the full frequency-domain QNM spectrum in the time-domain signal. Even then, the ``new'' frequencies associated with the cavity will be observable only if the amplitude of the perturbations is large enough to overcome the power-law tail.
Therefore, all calculations of QNM frequencies using modified potentials and/or modified boundary conditions should be complemented by time-domain studies to verify that these modifications affect the prompt ringdown, which is observationally accessible to interferometric detectors.

One last crucial remark is that this work only addressed the spectral instability of the {\em fundamental} QNM.
Spectral instabilities in the overtones due to short wavelength-(UV) perturbations could affect the time-domain signal even at early times, and they may be detectable~\cite{Jaramillo:2021tmt}.
The perturbative bumps considered in this work can more easily destabilize the higher overtones that would then be replaced by trapped-mode frequencies, and the time-domain waveform at late times would then contain modes from the ``new'' (destabilized) spectrum as long as they are not dominated in amplitude by the power-law tail.

Our main finding here is that the dominant mode of the prompt ringdown is very close in frequency to the fundamental mode of the unperturbed potential. However, our current fitting procedure does not allow us to conclude that higher overtones would also remain close to their unperturbed values. While we do not see any reason to expect that the instability of the rest of the spectrum behaves differently, the detection of overtones from the time-domain signal is not as straightforward as the detection of the fundamental mode~\cite{Isi:2021iql,Cotesta:2022pci}. In fact, the extraction of overtone frequencies by fitting numerical relativity waveforms (or even time evolution of the Regge-Wheeler and Zerilli equations) is nontrivial and often inaccurate. Any conclusions on the observational stability of higher overtones require a more careful investigation that we will address in future work.

\acknowledgments
We are thankful to Luis Lehner for useful correspondence, and to an anonymous reviewer for many useful comments which improved the overall quality of the manuscript.
E.B. and M.H.-Y.C. are supported by NSF Grants No. AST-2006538, No. PHY-2207502, No. PHY-090003, and No. PHY-20043, and NASA Grants No. 19-ATP19-0051, No. 20-LPS20-0011, and No. 21-ATP21-0010. This research project was conducted using computational resources at the Maryland Advanced Research Computing Center (MARCC). 
V.C. is a Villum Investigator and a DNRF Chair, supported by VILLUM FONDEN (grant no. 37766) and by the Danish Research Foundation. V.C. acknowledges financial support provided under the European
Union’s H2020 ERC Advanced Grant “Black holes: gravitational engines of discovery” grant agreement
no. Gravitas–101052587.
This project has received funding from the European Union's Horizon 2020 research and innovation programme under the Marie Sklodowska-Curie grant agreement No 101007855.
We thank FCT for financial support through Projects~No.~UIDB/00099/2020 and No. UIDB/04459/2020.
We acknowledge financial support provided by FCT/Portugal through grants No. PTDC/MAT-APL/30043/2017 and No. PTDC/FIS-AST/7002/2020.
FDF acknowledges financial support by Japan Society for the Promotion of Science Grants-in-Aid for international research fellow No. 21P21318. 
P.M. acknowledges support from the Japanese Government (MEXT) scholarship for Research Student. 
S.M.'s work was supported in part by JSPS Grants-in-Aid for Scientific Research No.~17H02890, No.~17H06359, and by World Premier International Research Center Initiative, MEXT, Japan. 

\bibliography{biblio}

\begin{thebibliography}{64}%
\makeatletter
\providecommand \@ifxundefined [1]{%
 \@ifx{#1\undefined}
}%
\providecommand \@ifnum [1]{%
 \ifnum #1\expandafter \@firstoftwo
 \else \expandafter \@secondoftwo
 \fi
}%
\providecommand \@ifx [1]{%
 \ifx #1\expandafter \@firstoftwo
 \else \expandafter \@secondoftwo
 \fi
}%
\providecommand \natexlab [1]{#1}%
\providecommand \enquote  [1]{``#1''}%
\providecommand \bibnamefont  [1]{#1}%
\providecommand \bibfnamefont [1]{#1}%
\providecommand \citenamefont [1]{#1}%
\providecommand \href@noop [0]{\@secondoftwo}%
\providecommand \href [0]{\begingroup \@sanitize@url \@href}%
\providecommand \@href[1]{\@@startlink{#1}\@@href}%
\providecommand \@@href[1]{\endgroup#1\@@endlink}%
\providecommand \@sanitize@url [0]{\catcode `\\12\catcode `\$12\catcode
  `\&12\catcode `\#12\catcode `\^12\catcode `\_12\catcode `\%12\relax}%
\providecommand \@@startlink[1]{}%
\providecommand \@@endlink[0]{}%
\providecommand \url  [0]{\begingroup\@sanitize@url \@url }%
\providecommand \@url [1]{\endgroup\@href {#1}{\urlprefix }}%
\providecommand \urlprefix  [0]{URL }%
\providecommand \Eprint [0]{\href }%
\providecommand \doibase [0]{http://dx.doi.org/}%
\providecommand \selectlanguage [0]{\@gobble}%
\providecommand \bibinfo  [0]{\@secondoftwo}%
\providecommand \bibfield  [0]{\@secondoftwo}%
\providecommand \translation [1]{[#1]}%
\providecommand \BibitemOpen [0]{}%
\providecommand \bibitemStop [0]{}%
\providecommand \bibitemNoStop [0]{.\EOS\space}%
\providecommand \EOS [0]{\spacefactor3000\relax}%
\providecommand \BibitemShut  [1]{\csname bibitem#1\endcsname}%
\let\auto@bib@innerbib\@empty
%</preamble>
\bibitem [{\citenamefont {Abbott}\ \emph
  {et~al.}(2016{\natexlab{a}})\citenamefont {Abbott} \emph
  {et~al.}}]{LIGO:2016}%
  \BibitemOpen
  \bibfield  {author} {\bibinfo {author} {\bibfnamefont {B.~P.}\ \bibnamefont
  {Abbott}} \emph {et~al.} (\bibinfo {collaboration} {LIGO Scientific,
  Virgo}),\ }\href {\doibase 10.1103/PhysRevLett.116.061102} {\bibfield
  {journal} {\bibinfo  {journal} {Phys. Rev. Lett.}\ }\textbf {\bibinfo
  {volume} {116}},\ \bibinfo {pages} {061102} (\bibinfo {year}
  {2016}{\natexlab{a}})},\ \Eprint {http://arxiv.org/abs/1602.03837}
  {arXiv:1602.03837 [gr-qc]} \BibitemShut {NoStop}%
\bibitem [{\citenamefont {Abbott}\ \emph {et~al.}(2021)\citenamefont {Abbott}
  \emph {et~al.}}]{LIGO:2020}%
  \BibitemOpen
  \bibfield  {author} {\bibinfo {author} {\bibfnamefont {R.}~\bibnamefont
  {Abbott}} \emph {et~al.} (\bibinfo {collaboration} {LIGO Scientific,
  Virgo}),\ }\href {\doibase 10.1103/PhysRevX.11.021053} {\bibfield  {journal}
  {\bibinfo  {journal} {Phys. Rev. X}\ }\textbf {\bibinfo {volume} {11}},\
  \bibinfo {pages} {021053} (\bibinfo {year} {2021})},\ \Eprint
  {http://arxiv.org/abs/2010.14527} {arXiv:2010.14527 [gr-qc]} \BibitemShut
  {NoStop}%
\bibitem [{\citenamefont {Akiyama}\ \emph {et~al.}(2019)\citenamefont {Akiyama}
  \emph {et~al.}}]{EHT:2019}%
  \BibitemOpen
  \bibfield  {author} {\bibinfo {author} {\bibfnamefont {K.}~\bibnamefont
  {Akiyama}} \emph {et~al.} (\bibinfo {collaboration} {Event Horizon
  Telescope}),\ }\href {\doibase 10.3847/2041-8213/ab0ec7} {\bibfield
  {journal} {\bibinfo  {journal} {Astrophys. J. Lett.}\ }\textbf {\bibinfo
  {volume} {875}},\ \bibinfo {pages} {L1} (\bibinfo {year} {2019})},\ \Eprint
  {http://arxiv.org/abs/1906.11238} {arXiv:1906.11238 [astro-ph.GA]}
  \BibitemShut {NoStop}%
\bibitem [{\citenamefont {Berti}\ \emph {et~al.}(2015)\citenamefont {Berti}
  \emph {et~al.}}]{Berti:2015itd}%
  \BibitemOpen
  \bibfield  {author} {\bibinfo {author} {\bibfnamefont {E.}~\bibnamefont
  {Berti}} \emph {et~al.},\ }\href {\doibase 10.1088/0264-9381/32/24/243001}
  {\bibfield  {journal} {\bibinfo  {journal} {Class. Quant. Grav.}\ }\textbf
  {\bibinfo {volume} {32}},\ \bibinfo {pages} {243001} (\bibinfo {year}
  {2015})},\ \Eprint {http://arxiv.org/abs/1501.07274} {arXiv:1501.07274
  [gr-qc]} \BibitemShut {NoStop}%
\bibitem [{\citenamefont {Cardoso}\ and\ \citenamefont
  {Pani}(2017)}]{Cardoso:2017cqb}%
  \BibitemOpen
  \bibfield  {author} {\bibinfo {author} {\bibfnamefont {V.}~\bibnamefont
  {Cardoso}}\ and\ \bibinfo {author} {\bibfnamefont {P.}~\bibnamefont {Pani}},\
  }\href {\doibase 10.1038/s41550-017-0225-y} {\bibfield  {journal} {\bibinfo
  {journal} {Nature Astron.}\ }\textbf {\bibinfo {volume} {1}},\ \bibinfo
  {pages} {586} (\bibinfo {year} {2017})},\ \Eprint
  {http://arxiv.org/abs/1709.01525} {arXiv:1709.01525 [gr-qc]} \BibitemShut
  {NoStop}%
\bibitem [{\citenamefont {Berti}\ \emph
  {et~al.}(2018{\natexlab{a}})\citenamefont {Berti}, \citenamefont {Yagi},\
  and\ \citenamefont {Yunes}}]{Berti:2018cxi}%
  \BibitemOpen
  \bibfield  {author} {\bibinfo {author} {\bibfnamefont {E.}~\bibnamefont
  {Berti}}, \bibinfo {author} {\bibfnamefont {K.}~\bibnamefont {Yagi}}, \ and\
  \bibinfo {author} {\bibfnamefont {N.}~\bibnamefont {Yunes}},\ }\href
  {\doibase 10.1007/s10714-018-2362-8} {\bibfield  {journal} {\bibinfo
  {journal} {Gen. Rel. Grav.}\ }\textbf {\bibinfo {volume} {50}},\ \bibinfo
  {pages} {46} (\bibinfo {year} {2018}{\natexlab{a}})},\ \Eprint
  {http://arxiv.org/abs/1801.03208} {arXiv:1801.03208 [gr-qc]} \BibitemShut
  {NoStop}%
\bibitem [{\citenamefont {Berti}\ \emph
  {et~al.}(2018{\natexlab{b}})\citenamefont {Berti}, \citenamefont {Yagi},
  \citenamefont {Yang},\ and\ \citenamefont {Yunes}}]{Berti:2018vdi}%
  \BibitemOpen
  \bibfield  {author} {\bibinfo {author} {\bibfnamefont {E.}~\bibnamefont
  {Berti}}, \bibinfo {author} {\bibfnamefont {K.}~\bibnamefont {Yagi}},
  \bibinfo {author} {\bibfnamefont {H.}~\bibnamefont {Yang}}, \ and\ \bibinfo
  {author} {\bibfnamefont {N.}~\bibnamefont {Yunes}},\ }\href {\doibase
  10.1007/s10714-018-2372-6} {\bibfield  {journal} {\bibinfo  {journal} {Gen.
  Rel. Grav.}\ }\textbf {\bibinfo {volume} {50}},\ \bibinfo {pages} {49}
  (\bibinfo {year} {2018}{\natexlab{b}})},\ \Eprint
  {http://arxiv.org/abs/1801.03587} {arXiv:1801.03587 [gr-qc]} \BibitemShut
  {NoStop}%
\bibitem [{\citenamefont {Barack}\ \emph {et~al.}(2019)\citenamefont {Barack}
  \emph {et~al.}}]{Barack:2018yly}%
  \BibitemOpen
  \bibfield  {author} {\bibinfo {author} {\bibfnamefont {L.}~\bibnamefont
  {Barack}} \emph {et~al.},\ }\href {\doibase 10.1088/1361-6382/ab0587}
  {\bibfield  {journal} {\bibinfo  {journal} {Class. Quant. Grav.}\ }\textbf
  {\bibinfo {volume} {36}},\ \bibinfo {pages} {143001} (\bibinfo {year}
  {2019})},\ \Eprint {http://arxiv.org/abs/1806.05195} {arXiv:1806.05195
  [gr-qc]} \BibitemShut {NoStop}%
\bibitem [{\citenamefont {Carballo-Rubio}\ \emph {et~al.}(2018)\citenamefont
  {Carballo-Rubio}, \citenamefont {Di~Filippo}, \citenamefont {Liberati},\ and\
  \citenamefont {Visser}}]{Carballo:2018}%
  \BibitemOpen
  \bibfield  {author} {\bibinfo {author} {\bibfnamefont {R.}~\bibnamefont
  {Carballo-Rubio}}, \bibinfo {author} {\bibfnamefont {F.}~\bibnamefont
  {Di~Filippo}}, \bibinfo {author} {\bibfnamefont {S.}~\bibnamefont
  {Liberati}}, \ and\ \bibinfo {author} {\bibfnamefont {M.}~\bibnamefont
  {Visser}},\ }\href {\doibase 10.1103/PhysRevD.98.124009} {\bibfield
  {journal} {\bibinfo  {journal} {Phys. Rev. D}\ }\textbf {\bibinfo {volume}
  {98}},\ \bibinfo {pages} {124009} (\bibinfo {year} {2018})},\ \Eprint
  {http://arxiv.org/abs/1809.08238} {arXiv:1809.08238 [gr-qc]} \BibitemShut
  {NoStop}%
\bibitem [{\citenamefont {Cardoso}\ and\ \citenamefont
  {Pani}(2019)}]{Cardoso:2019rvt}%
  \BibitemOpen
  \bibfield  {author} {\bibinfo {author} {\bibfnamefont {V.}~\bibnamefont
  {Cardoso}}\ and\ \bibinfo {author} {\bibfnamefont {P.}~\bibnamefont {Pani}},\
  }\href {\doibase 10.1007/s41114-019-0020-4} {\bibfield  {journal} {\bibinfo
  {journal} {Living Rev. Rel.}\ }\textbf {\bibinfo {volume} {22}},\ \bibinfo
  {pages} {4} (\bibinfo {year} {2019})},\ \Eprint
  {http://arxiv.org/abs/1904.05363} {arXiv:1904.05363 [gr-qc]} \BibitemShut
  {NoStop}%
\bibitem [{\citenamefont {Detweiler}(1980)}]{Detweiler:1980gk}%
  \BibitemOpen
  \bibfield  {author} {\bibinfo {author} {\bibfnamefont {S.~L.}\ \bibnamefont
  {Detweiler}},\ }\href {\doibase 10.1086/158109} {\bibfield  {journal}
  {\bibinfo  {journal} {Astrophys. J.}\ }\textbf {\bibinfo {volume} {239}},\
  \bibinfo {pages} {292} (\bibinfo {year} {1980})}\BibitemShut {NoStop}%
\bibitem [{\citenamefont {Dreyer}\ \emph {et~al.}(2004)\citenamefont {Dreyer},
  \citenamefont {Kelly}, \citenamefont {Krishnan}, \citenamefont {Finn},
  \citenamefont {Garrison},\ and\ \citenamefont
  {Lopez-Aleman}}]{Dreyer:2003bv}%
  \BibitemOpen
  \bibfield  {author} {\bibinfo {author} {\bibfnamefont {O.}~\bibnamefont
  {Dreyer}}, \bibinfo {author} {\bibfnamefont {B.~J.}\ \bibnamefont {Kelly}},
  \bibinfo {author} {\bibfnamefont {B.}~\bibnamefont {Krishnan}}, \bibinfo
  {author} {\bibfnamefont {L.~S.}\ \bibnamefont {Finn}}, \bibinfo {author}
  {\bibfnamefont {D.}~\bibnamefont {Garrison}}, \ and\ \bibinfo {author}
  {\bibfnamefont {R.}~\bibnamefont {Lopez-Aleman}},\ }\href {\doibase
  10.1088/0264-9381/21/4/003} {\bibfield  {journal} {\bibinfo  {journal}
  {Class. Quant. Grav.}\ }\textbf {\bibinfo {volume} {21}},\ \bibinfo {pages}
  {787} (\bibinfo {year} {2004})},\ \Eprint
  {http://arxiv.org/abs/gr-qc/0309007} {arXiv:gr-qc/0309007} \BibitemShut
  {NoStop}%
\bibitem [{\citenamefont {Berti}\ \emph {et~al.}(2006)\citenamefont {Berti},
  \citenamefont {Cardoso},\ and\ \citenamefont {Will}}]{Berti:2005ys}%
  \BibitemOpen
  \bibfield  {author} {\bibinfo {author} {\bibfnamefont {E.}~\bibnamefont
  {Berti}}, \bibinfo {author} {\bibfnamefont {V.}~\bibnamefont {Cardoso}}, \
  and\ \bibinfo {author} {\bibfnamefont {C.~M.}\ \bibnamefont {Will}},\ }\href
  {\doibase 10.1103/PhysRevD.73.064030} {\bibfield  {journal} {\bibinfo
  {journal} {Phys. Rev. D}\ }\textbf {\bibinfo {volume} {73}},\ \bibinfo
  {pages} {064030} (\bibinfo {year} {2006})},\ \Eprint
  {http://arxiv.org/abs/gr-qc/0512160} {arXiv:gr-qc/0512160} \BibitemShut
  {NoStop}%
\bibitem [{\citenamefont {Abbott}\ \emph
  {et~al.}(2016{\natexlab{b}})\citenamefont {Abbott} \emph
  {et~al.}}]{LIGO:2016lio}%
  \BibitemOpen
  \bibfield  {author} {\bibinfo {author} {\bibfnamefont {B.~P.}\ \bibnamefont
  {Abbott}} \emph {et~al.} (\bibinfo {collaboration} {LIGO Scientific,
  Virgo}),\ }\href {\doibase 10.1103/PhysRevLett.116.221101} {\bibfield
  {journal} {\bibinfo  {journal} {Phys. Rev. Lett.}\ }\textbf {\bibinfo
  {volume} {116}},\ \bibinfo {pages} {221101} (\bibinfo {year}
  {2016}{\natexlab{b}})},\ \bibinfo {note} {[Erratum: Phys.Rev.Lett. 121,
  129902 (2018)]},\ \Eprint {http://arxiv.org/abs/1602.03841} {arXiv:1602.03841
  [gr-qc]} \BibitemShut {NoStop}%
\bibitem [{\citenamefont {Kokkotas}\ and\ \citenamefont
  {Schmidt}(1999)}]{Kokkotas:1999bd}%
  \BibitemOpen
  \bibfield  {author} {\bibinfo {author} {\bibfnamefont {K.~D.}\ \bibnamefont
  {Kokkotas}}\ and\ \bibinfo {author} {\bibfnamefont {B.~G.}\ \bibnamefont
  {Schmidt}},\ }\href {\doibase 10.12942/lrr-1999-2} {\bibfield  {journal}
  {\bibinfo  {journal} {Living Rev. Rel.}\ }\textbf {\bibinfo {volume} {2}},\
  \bibinfo {pages} {2} (\bibinfo {year} {1999})},\ \Eprint
  {http://arxiv.org/abs/gr-qc/9909058} {arXiv:gr-qc/9909058} \BibitemShut
  {NoStop}%
\bibitem [{\citenamefont {Berti}\ \emph {et~al.}(2009)\citenamefont {Berti},
  \citenamefont {Cardoso},\ and\ \citenamefont {Starinets}}]{Berti:2009kk}%
  \BibitemOpen
  \bibfield  {author} {\bibinfo {author} {\bibfnamefont {E.}~\bibnamefont
  {Berti}}, \bibinfo {author} {\bibfnamefont {V.}~\bibnamefont {Cardoso}}, \
  and\ \bibinfo {author} {\bibfnamefont {A.~O.}\ \bibnamefont {Starinets}},\
  }\href {\doibase 10.1088/0264-9381/26/16/163001} {\bibfield  {journal}
  {\bibinfo  {journal} {Class. Quant. Grav.}\ }\textbf {\bibinfo {volume}
  {26}},\ \bibinfo {pages} {163001} (\bibinfo {year} {2009})},\ \Eprint
  {http://arxiv.org/abs/0905.2975} {arXiv:0905.2975 [gr-qc]} \BibitemShut
  {NoStop}%
\bibitem [{\citenamefont {Jaramillo}\ \emph
  {et~al.}(2021{\natexlab{a}})\citenamefont {Jaramillo}, \citenamefont
  {Panosso~Macedo},\ and\ \citenamefont {Al~Sheikh}}]{Jaramillo:2020tuu}%
  \BibitemOpen
  \bibfield  {author} {\bibinfo {author} {\bibfnamefont {J.~L.}\ \bibnamefont
  {Jaramillo}}, \bibinfo {author} {\bibfnamefont {R.}~\bibnamefont
  {Panosso~Macedo}}, \ and\ \bibinfo {author} {\bibfnamefont {L.}~\bibnamefont
  {Al~Sheikh}},\ }\href {\doibase 10.1103/PhysRevX.11.031003} {\bibfield
  {journal} {\bibinfo  {journal} {Phys. Rev. X}\ }\textbf {\bibinfo {volume}
  {11}},\ \bibinfo {pages} {031003} (\bibinfo {year} {2021}{\natexlab{a}})},\
  \Eprint {http://arxiv.org/abs/2004.06434} {arXiv:2004.06434 [gr-qc]}
  \BibitemShut {NoStop}%
\bibitem [{\citenamefont {Jaramillo}\ \emph
  {et~al.}(2021{\natexlab{b}})\citenamefont {Jaramillo}, \citenamefont
  {Panosso~Macedo},\ and\ \citenamefont {Sheikh}}]{Jaramillo:2021tmt}%
  \BibitemOpen
  \bibfield  {author} {\bibinfo {author} {\bibfnamefont {J.~L.}\ \bibnamefont
  {Jaramillo}}, \bibinfo {author} {\bibfnamefont {R.}~\bibnamefont
  {Panosso~Macedo}}, \ and\ \bibinfo {author} {\bibfnamefont {L.~A.}\
  \bibnamefont {Sheikh}},\ }\href@noop {} {\  (\bibinfo {year}
  {2021}{\natexlab{b}})},\ \Eprint {http://arxiv.org/abs/2105.03451}
  {arXiv:2105.03451 [gr-qc]} \BibitemShut {NoStop}%
\bibitem [{\citenamefont {Destounis}\ \emph {et~al.}(2021)\citenamefont
  {Destounis}, \citenamefont {Macedo}, \citenamefont {Berti}, \citenamefont
  {Cardoso},\ and\ \citenamefont {Jaramillo}}]{Destounis:2021lum}%
  \BibitemOpen
  \bibfield  {author} {\bibinfo {author} {\bibfnamefont {K.}~\bibnamefont
  {Destounis}}, \bibinfo {author} {\bibfnamefont {R.~P.}\ \bibnamefont
  {Macedo}}, \bibinfo {author} {\bibfnamefont {E.}~\bibnamefont {Berti}},
  \bibinfo {author} {\bibfnamefont {V.}~\bibnamefont {Cardoso}}, \ and\
  \bibinfo {author} {\bibfnamefont {J.~L.}\ \bibnamefont {Jaramillo}},\ }\href
  {\doibase 10.1103/PhysRevD.104.084091} {\bibfield  {journal} {\bibinfo
  {journal} {Phys. Rev. D}\ }\textbf {\bibinfo {volume} {104}},\ \bibinfo
  {pages} {084091} (\bibinfo {year} {2021})},\ \Eprint
  {http://arxiv.org/abs/2107.09673} {arXiv:2107.09673 [gr-qc]} \BibitemShut
  {NoStop}%
\bibitem [{\citenamefont {Gasperin}\ and\ \citenamefont
  {Jaramillo}(2021)}]{Gasperin:2021kfv}%
  \BibitemOpen
  \bibfield  {author} {\bibinfo {author} {\bibfnamefont {E.}~\bibnamefont
  {Gasperin}}\ and\ \bibinfo {author} {\bibfnamefont {J.~L.}\ \bibnamefont
  {Jaramillo}},\ }\href@noop {} {\  (\bibinfo {year} {2021})},\ \Eprint
  {http://arxiv.org/abs/2107.12865} {arXiv:2107.12865 [gr-qc]} \BibitemShut
  {NoStop}%
\bibitem [{\citenamefont {Nollert}(1996)}]{Nollert:1996rf}%
  \BibitemOpen
  \bibfield  {author} {\bibinfo {author} {\bibfnamefont {H.-P.}\ \bibnamefont
  {Nollert}},\ }\href {\doibase 10.1103/PhysRevD.53.4397} {\bibfield  {journal}
  {\bibinfo  {journal} {Phys. Rev. D}\ }\textbf {\bibinfo {volume} {53}},\
  \bibinfo {pages} {4397} (\bibinfo {year} {1996})},\ \Eprint
  {http://arxiv.org/abs/gr-qc/9602032} {arXiv:gr-qc/9602032} \BibitemShut
  {NoStop}%
\bibitem [{\citenamefont {Cheung}\ \emph {et~al.}(2022)\citenamefont {Cheung},
  \citenamefont {Destounis}, \citenamefont {Macedo}, \citenamefont {Berti},\
  and\ \citenamefont {Cardoso}}]{Cheung:2021bol}%
  \BibitemOpen
  \bibfield  {author} {\bibinfo {author} {\bibfnamefont {M.~H.-Y.}\
  \bibnamefont {Cheung}}, \bibinfo {author} {\bibfnamefont {K.}~\bibnamefont
  {Destounis}}, \bibinfo {author} {\bibfnamefont {R.~P.}\ \bibnamefont
  {Macedo}}, \bibinfo {author} {\bibfnamefont {E.}~\bibnamefont {Berti}}, \
  and\ \bibinfo {author} {\bibfnamefont {V.}~\bibnamefont {Cardoso}},\ }\href
  {\doibase 10.1103/PhysRevLett.128.111103} {\bibfield  {journal} {\bibinfo
  {journal} {Phys. Rev. Lett.}\ }\textbf {\bibinfo {volume} {128}},\ \bibinfo
  {pages} {111103} (\bibinfo {year} {2022})},\ \Eprint
  {http://arxiv.org/abs/2111.05415} {arXiv:2111.05415 [gr-qc]} \BibitemShut
  {NoStop}%
\bibitem [{\citenamefont {Gleiser}\ \emph {et~al.}(2000)\citenamefont
  {Gleiser}, \citenamefont {Nicasio}, \citenamefont {Price},\ and\
  \citenamefont {Pullin}}]{Gleiser:1998rw}%
  \BibitemOpen
  \bibfield  {author} {\bibinfo {author} {\bibfnamefont {R.~J.}\ \bibnamefont
  {Gleiser}}, \bibinfo {author} {\bibfnamefont {C.~O.}\ \bibnamefont
  {Nicasio}}, \bibinfo {author} {\bibfnamefont {R.~H.}\ \bibnamefont {Price}},
  \ and\ \bibinfo {author} {\bibfnamefont {J.}~\bibnamefont {Pullin}},\ }\href
  {\doibase 10.1016/S0370-1573(99)00048-4} {\bibfield  {journal} {\bibinfo
  {journal} {Phys. Rept.}\ }\textbf {\bibinfo {volume} {325}},\ \bibinfo
  {pages} {41} (\bibinfo {year} {2000})},\ \Eprint
  {http://arxiv.org/abs/gr-qc/9807077} {arXiv:gr-qc/9807077} \BibitemShut
  {NoStop}%
\bibitem [{\citenamefont {Campanelli}\ and\ \citenamefont
  {Lousto}(1999)}]{Campanelli:1998jv}%
  \BibitemOpen
  \bibfield  {author} {\bibinfo {author} {\bibfnamefont {M.}~\bibnamefont
  {Campanelli}}\ and\ \bibinfo {author} {\bibfnamefont {C.~O.}\ \bibnamefont
  {Lousto}},\ }\href {\doibase 10.1103/PhysRevD.59.124022} {\bibfield
  {journal} {\bibinfo  {journal} {Phys. Rev. D}\ }\textbf {\bibinfo {volume}
  {59}},\ \bibinfo {pages} {124022} (\bibinfo {year} {1999})},\ \Eprint
  {http://arxiv.org/abs/gr-qc/9811019} {arXiv:gr-qc/9811019} \BibitemShut
  {NoStop}%
\bibitem [{\citenamefont {Zlochower}\ \emph {et~al.}(2003)\citenamefont
  {Zlochower}, \citenamefont {Gomez}, \citenamefont {Husa}, \citenamefont
  {Lehner},\ and\ \citenamefont {Winicour}}]{Zlochower:2003yh}%
  \BibitemOpen
  \bibfield  {author} {\bibinfo {author} {\bibfnamefont {Y.}~\bibnamefont
  {Zlochower}}, \bibinfo {author} {\bibfnamefont {R.}~\bibnamefont {Gomez}},
  \bibinfo {author} {\bibfnamefont {S.}~\bibnamefont {Husa}}, \bibinfo {author}
  {\bibfnamefont {L.}~\bibnamefont {Lehner}}, \ and\ \bibinfo {author}
  {\bibfnamefont {J.}~\bibnamefont {Winicour}},\ }\href {\doibase
  10.1103/PhysRevD.68.084014} {\bibfield  {journal} {\bibinfo  {journal} {Phys.
  Rev. D}\ }\textbf {\bibinfo {volume} {68}},\ \bibinfo {pages} {084014}
  (\bibinfo {year} {2003})},\ \Eprint {http://arxiv.org/abs/gr-qc/0306098}
  {arXiv:gr-qc/0306098} \BibitemShut {NoStop}%
\bibitem [{\citenamefont {Nakano}\ and\ \citenamefont
  {Ioka}(2007)}]{Nakano:2007cj}%
  \BibitemOpen
  \bibfield  {author} {\bibinfo {author} {\bibfnamefont {H.}~\bibnamefont
  {Nakano}}\ and\ \bibinfo {author} {\bibfnamefont {K.}~\bibnamefont {Ioka}},\
  }\href {\doibase 10.1103/PhysRevD.76.084007} {\bibfield  {journal} {\bibinfo
  {journal} {Phys. Rev. D}\ }\textbf {\bibinfo {volume} {76}},\ \bibinfo
  {pages} {084007} (\bibinfo {year} {2007})},\ \Eprint
  {http://arxiv.org/abs/0708.0450} {arXiv:0708.0450 [gr-qc]} \BibitemShut
  {NoStop}%
\bibitem [{\citenamefont {Ioka}\ and\ \citenamefont
  {Nakano}(2007)}]{Ioka:2007ak}%
  \BibitemOpen
  \bibfield  {author} {\bibinfo {author} {\bibfnamefont {K.}~\bibnamefont
  {Ioka}}\ and\ \bibinfo {author} {\bibfnamefont {H.}~\bibnamefont {Nakano}},\
  }\href {\doibase 10.1103/PhysRevD.76.061503} {\bibfield  {journal} {\bibinfo
  {journal} {Phys. Rev. D}\ }\textbf {\bibinfo {volume} {76}},\ \bibinfo
  {pages} {061503} (\bibinfo {year} {2007})},\ \Eprint
  {http://arxiv.org/abs/0704.3467} {arXiv:0704.3467 [astro-ph]} \BibitemShut
  {NoStop}%
\bibitem [{\citenamefont {Okuzumi}\ \emph {et~al.}(2008)\citenamefont
  {Okuzumi}, \citenamefont {Ioka},\ and\ \citenamefont
  {Sakagami}}]{Okuzumi:2008ej}%
  \BibitemOpen
  \bibfield  {author} {\bibinfo {author} {\bibfnamefont {S.}~\bibnamefont
  {Okuzumi}}, \bibinfo {author} {\bibfnamefont {K.}~\bibnamefont {Ioka}}, \
  and\ \bibinfo {author} {\bibfnamefont {M.-a.}\ \bibnamefont {Sakagami}},\
  }\href {\doibase 10.1103/PhysRevD.77.124018} {\bibfield  {journal} {\bibinfo
  {journal} {Phys. Rev. D}\ }\textbf {\bibinfo {volume} {77}},\ \bibinfo
  {pages} {124018} (\bibinfo {year} {2008})},\ \Eprint
  {http://arxiv.org/abs/0803.0501} {arXiv:0803.0501 [gr-qc]} \BibitemShut
  {NoStop}%
\bibitem [{\citenamefont {Pazos}\ \emph {et~al.}(2010)\citenamefont {Pazos},
  \citenamefont {Brizuela}, \citenamefont {Martin-Garcia},\ and\ \citenamefont
  {Tiglio}}]{Pazos:2010xf}%
  \BibitemOpen
  \bibfield  {author} {\bibinfo {author} {\bibfnamefont {E.}~\bibnamefont
  {Pazos}}, \bibinfo {author} {\bibfnamefont {D.}~\bibnamefont {Brizuela}},
  \bibinfo {author} {\bibfnamefont {J.~M.}\ \bibnamefont {Martin-Garcia}}, \
  and\ \bibinfo {author} {\bibfnamefont {M.}~\bibnamefont {Tiglio}},\ }\href
  {\doibase 10.1103/PhysRevD.82.104028} {\bibfield  {journal} {\bibinfo
  {journal} {Phys. Rev. D}\ }\textbf {\bibinfo {volume} {82}},\ \bibinfo
  {pages} {104028} (\bibinfo {year} {2010})},\ \Eprint
  {http://arxiv.org/abs/1009.4665} {arXiv:1009.4665 [gr-qc]} \BibitemShut
  {NoStop}%
\bibitem [{\citenamefont {Sberna}\ \emph {et~al.}(2022)\citenamefont {Sberna},
  \citenamefont {Bosch}, \citenamefont {East}, \citenamefont {Green},\ and\
  \citenamefont {Lehner}}]{Sberna:2021eui}%
  \BibitemOpen
  \bibfield  {author} {\bibinfo {author} {\bibfnamefont {L.}~\bibnamefont
  {Sberna}}, \bibinfo {author} {\bibfnamefont {P.}~\bibnamefont {Bosch}},
  \bibinfo {author} {\bibfnamefont {W.~E.}\ \bibnamefont {East}}, \bibinfo
  {author} {\bibfnamefont {S.~R.}\ \bibnamefont {Green}}, \ and\ \bibinfo
  {author} {\bibfnamefont {L.}~\bibnamefont {Lehner}},\ }\href {\doibase
  10.1103/PhysRevD.105.064046} {\bibfield  {journal} {\bibinfo  {journal}
  {Phys. Rev. D}\ }\textbf {\bibinfo {volume} {105}},\ \bibinfo {pages}
  {064046} (\bibinfo {year} {2022})},\ \Eprint
  {http://arxiv.org/abs/2112.11168} {arXiv:2112.11168 [gr-qc]} \BibitemShut
  {NoStop}%
\bibitem [{\citenamefont {Leung}\ \emph {et~al.}(1997)\citenamefont {Leung},
  \citenamefont {Liu}, \citenamefont {Suen}, \citenamefont {Tam},\ and\
  \citenamefont {Young}}]{Leung:1997was}%
  \BibitemOpen
  \bibfield  {author} {\bibinfo {author} {\bibfnamefont {P.~T.}\ \bibnamefont
  {Leung}}, \bibinfo {author} {\bibfnamefont {Y.~T.}\ \bibnamefont {Liu}},
  \bibinfo {author} {\bibfnamefont {W.~M.}\ \bibnamefont {Suen}}, \bibinfo
  {author} {\bibfnamefont {C.~Y.}\ \bibnamefont {Tam}}, \ and\ \bibinfo
  {author} {\bibfnamefont {K.}~\bibnamefont {Young}},\ }\href {\doibase
  10.1103/PhysRevLett.78.2894} {\bibfield  {journal} {\bibinfo  {journal}
  {Phys. Rev. Lett.}\ }\textbf {\bibinfo {volume} {78}},\ \bibinfo {pages}
  {2894} (\bibinfo {year} {1997})},\ \Eprint
  {http://arxiv.org/abs/gr-qc/9903031} {arXiv:gr-qc/9903031} \BibitemShut
  {NoStop}%
\bibitem [{\citenamefont {Leung}\ \emph {et~al.}(1999)\citenamefont {Leung},
  \citenamefont {Liu}, \citenamefont {Suen}, \citenamefont {Tam},\ and\
  \citenamefont {Young}}]{Leung:1999iq}%
  \BibitemOpen
  \bibfield  {author} {\bibinfo {author} {\bibfnamefont {P.~T.}\ \bibnamefont
  {Leung}}, \bibinfo {author} {\bibfnamefont {Y.~T.}\ \bibnamefont {Liu}},
  \bibinfo {author} {\bibfnamefont {W.~M.}\ \bibnamefont {Suen}}, \bibinfo
  {author} {\bibfnamefont {C.~Y.}\ \bibnamefont {Tam}}, \ and\ \bibinfo
  {author} {\bibfnamefont {K.}~\bibnamefont {Young}},\ }\href {\doibase
  10.1103/PhysRevD.59.044034} {\bibfield  {journal} {\bibinfo  {journal} {Phys.
  Rev. D}\ }\textbf {\bibinfo {volume} {59}},\ \bibinfo {pages} {044034}
  (\bibinfo {year} {1999})},\ \Eprint {http://arxiv.org/abs/gr-qc/9903032}
  {arXiv:gr-qc/9903032} \BibitemShut {NoStop}%
\bibitem [{\citenamefont {Barausse}\ \emph {et~al.}(2014)\citenamefont
  {Barausse}, \citenamefont {Cardoso},\ and\ \citenamefont
  {Pani}}]{Barausse:2014tra}%
  \BibitemOpen
  \bibfield  {author} {\bibinfo {author} {\bibfnamefont {E.}~\bibnamefont
  {Barausse}}, \bibinfo {author} {\bibfnamefont {V.}~\bibnamefont {Cardoso}}, \
  and\ \bibinfo {author} {\bibfnamefont {P.}~\bibnamefont {Pani}},\ }\href
  {\doibase 10.1103/PhysRevD.89.104059} {\bibfield  {journal} {\bibinfo
  {journal} {Phys. Rev. D}\ }\textbf {\bibinfo {volume} {89}},\ \bibinfo
  {pages} {104059} (\bibinfo {year} {2014})},\ \Eprint
  {http://arxiv.org/abs/1404.7149} {arXiv:1404.7149 [gr-qc]} \BibitemShut
  {NoStop}%
\bibitem [{\citenamefont {Chung}\ \emph {et~al.}(2021)\citenamefont {Chung},
  \citenamefont {Gais}, \citenamefont {Cheung},\ and\ \citenamefont
  {Li}}]{Chung:2021roh}%
  \BibitemOpen
  \bibfield  {author} {\bibinfo {author} {\bibfnamefont {A.~K.-W.}\
  \bibnamefont {Chung}}, \bibinfo {author} {\bibfnamefont {J.}~\bibnamefont
  {Gais}}, \bibinfo {author} {\bibfnamefont {M.~H.-Y.}\ \bibnamefont {Cheung}},
  \ and\ \bibinfo {author} {\bibfnamefont {T.~G.~F.}\ \bibnamefont {Li}},\
  }\href {\doibase 10.1103/PhysRevD.104.084028} {\bibfield  {journal} {\bibinfo
   {journal} {Phys. Rev. D}\ }\textbf {\bibinfo {volume} {104}},\ \bibinfo
  {pages} {084028} (\bibinfo {year} {2021})},\ \Eprint
  {http://arxiv.org/abs/2107.05492} {arXiv:2107.05492 [gr-qc]} \BibitemShut
  {NoStop}%
\bibitem [{\citenamefont {Cardoso}\ \emph
  {et~al.}(2019{\natexlab{a}})\citenamefont {Cardoso}, \citenamefont {Kimura},
  \citenamefont {Maselli}, \citenamefont {Berti}, \citenamefont {Macedo},\ and\
  \citenamefont {McManus}}]{Cardoso:2019mqo}%
  \BibitemOpen
  \bibfield  {author} {\bibinfo {author} {\bibfnamefont {V.}~\bibnamefont
  {Cardoso}}, \bibinfo {author} {\bibfnamefont {M.}~\bibnamefont {Kimura}},
  \bibinfo {author} {\bibfnamefont {A.}~\bibnamefont {Maselli}}, \bibinfo
  {author} {\bibfnamefont {E.}~\bibnamefont {Berti}}, \bibinfo {author}
  {\bibfnamefont {C.~F.~B.}\ \bibnamefont {Macedo}}, \ and\ \bibinfo {author}
  {\bibfnamefont {R.}~\bibnamefont {McManus}},\ }\href {\doibase
  10.1103/PhysRevD.99.104077} {\bibfield  {journal} {\bibinfo  {journal} {Phys.
  Rev. D}\ }\textbf {\bibinfo {volume} {99}},\ \bibinfo {pages} {104077}
  (\bibinfo {year} {2019}{\natexlab{a}})},\ \Eprint
  {http://arxiv.org/abs/1901.01265} {arXiv:1901.01265 [gr-qc]} \BibitemShut
  {NoStop}%
\bibitem [{\citenamefont {McManus}\ \emph {et~al.}(2019)\citenamefont
  {McManus}, \citenamefont {Berti}, \citenamefont {Macedo}, \citenamefont
  {Kimura}, \citenamefont {Maselli},\ and\ \citenamefont
  {Cardoso}}]{McManus:2019ulj}%
  \BibitemOpen
  \bibfield  {author} {\bibinfo {author} {\bibfnamefont {R.}~\bibnamefont
  {McManus}}, \bibinfo {author} {\bibfnamefont {E.}~\bibnamefont {Berti}},
  \bibinfo {author} {\bibfnamefont {C.~F.~B.}\ \bibnamefont {Macedo}}, \bibinfo
  {author} {\bibfnamefont {M.}~\bibnamefont {Kimura}}, \bibinfo {author}
  {\bibfnamefont {A.}~\bibnamefont {Maselli}}, \ and\ \bibinfo {author}
  {\bibfnamefont {V.}~\bibnamefont {Cardoso}},\ }\href {\doibase
  10.1103/PhysRevD.100.044061} {\bibfield  {journal} {\bibinfo  {journal}
  {Phys. Rev. D}\ }\textbf {\bibinfo {volume} {100}},\ \bibinfo {pages}
  {044061} (\bibinfo {year} {2019})},\ \Eprint
  {http://arxiv.org/abs/1906.05155} {arXiv:1906.05155 [gr-qc]} \BibitemShut
  {NoStop}%
\bibitem [{\citenamefont {Daghigh}\ \emph {et~al.}(2020)\citenamefont
  {Daghigh}, \citenamefont {Green},\ and\ \citenamefont
  {Morey}}]{Daghigh:2020jyk}%
  \BibitemOpen
  \bibfield  {author} {\bibinfo {author} {\bibfnamefont {R.~G.}\ \bibnamefont
  {Daghigh}}, \bibinfo {author} {\bibfnamefont {M.~D.}\ \bibnamefont {Green}},
  \ and\ \bibinfo {author} {\bibfnamefont {J.~C.}\ \bibnamefont {Morey}},\
  }\href {\doibase 10.1103/PhysRevD.101.104009} {\bibfield  {journal} {\bibinfo
   {journal} {Phys. Rev. D}\ }\textbf {\bibinfo {volume} {101}},\ \bibinfo
  {pages} {104009} (\bibinfo {year} {2020})},\ \Eprint
  {http://arxiv.org/abs/2002.07251} {arXiv:2002.07251 [gr-qc]} \BibitemShut
  {NoStop}%
\bibitem [{\citenamefont {Cardoso}\ \emph
  {et~al.}(2016{\natexlab{a}})\citenamefont {Cardoso}, \citenamefont
  {Franzin},\ and\ \citenamefont {Pani}}]{Cardoso:2016rao}%
  \BibitemOpen
  \bibfield  {author} {\bibinfo {author} {\bibfnamefont {V.}~\bibnamefont
  {Cardoso}}, \bibinfo {author} {\bibfnamefont {E.}~\bibnamefont {Franzin}}, \
  and\ \bibinfo {author} {\bibfnamefont {P.}~\bibnamefont {Pani}},\ }\href
  {\doibase 10.1103/PhysRevLett.116.171101} {\bibfield  {journal} {\bibinfo
  {journal} {Phys. Rev. Lett.}\ }\textbf {\bibinfo {volume} {116}},\ \bibinfo
  {pages} {171101} (\bibinfo {year} {2016}{\natexlab{a}})},\ \bibinfo {note}
  {[Erratum: Phys.Rev.Lett. 117, 089902 (2016)]},\ \Eprint
  {http://arxiv.org/abs/1602.07309} {arXiv:1602.07309 [gr-qc]} \BibitemShut
  {NoStop}%
\bibitem [{\citenamefont {Mark}\ \emph {et~al.}(2017)\citenamefont {Mark},
  \citenamefont {Zimmerman}, \citenamefont {Du},\ and\ \citenamefont
  {Chen}}]{Mark:2017dnq}%
  \BibitemOpen
  \bibfield  {author} {\bibinfo {author} {\bibfnamefont {Z.}~\bibnamefont
  {Mark}}, \bibinfo {author} {\bibfnamefont {A.}~\bibnamefont {Zimmerman}},
  \bibinfo {author} {\bibfnamefont {S.~M.}\ \bibnamefont {Du}}, \ and\ \bibinfo
  {author} {\bibfnamefont {Y.}~\bibnamefont {Chen}},\ }\href {\doibase
  10.1103/PhysRevD.96.084002} {\bibfield  {journal} {\bibinfo  {journal} {Phys.
  Rev. D}\ }\textbf {\bibinfo {volume} {96}},\ \bibinfo {pages} {084002}
  (\bibinfo {year} {2017})},\ \Eprint {http://arxiv.org/abs/1706.06155}
  {arXiv:1706.06155 [gr-qc]} \BibitemShut {NoStop}%
\bibitem [{\citenamefont {Hui}\ \emph {et~al.}(2019)\citenamefont {Hui},
  \citenamefont {Kabat},\ and\ \citenamefont {Wong}}]{Hui:2019aox}%
  \BibitemOpen
  \bibfield  {author} {\bibinfo {author} {\bibfnamefont {L.}~\bibnamefont
  {Hui}}, \bibinfo {author} {\bibfnamefont {D.}~\bibnamefont {Kabat}}, \ and\
  \bibinfo {author} {\bibfnamefont {S.~S.~C.}\ \bibnamefont {Wong}},\ }\href
  {\doibase 10.1088/1475-7516/2019/12/020} {\bibfield  {journal} {\bibinfo
  {journal} {JCAP}\ }\textbf {\bibinfo {volume} {12}},\ \bibinfo {pages} {020}
  (\bibinfo {year} {2019})},\ \Eprint {http://arxiv.org/abs/1909.10382}
  {arXiv:1909.10382 [gr-qc]} \BibitemShut {NoStop}%
\bibitem [{\citenamefont {Cardoso}\ \emph
  {et~al.}(2019{\natexlab{b}})\citenamefont {Cardoso}, \citenamefont {del
  Rio},\ and\ \citenamefont {Kimura}}]{Cardoso:2019nis}%
  \BibitemOpen
  \bibfield  {author} {\bibinfo {author} {\bibfnamefont {V.}~\bibnamefont
  {Cardoso}}, \bibinfo {author} {\bibfnamefont {A.}~\bibnamefont {del Rio}}, \
  and\ \bibinfo {author} {\bibfnamefont {M.}~\bibnamefont {Kimura}},\ }\href
  {\doibase 10.1103/PhysRevD.100.084046} {\bibfield  {journal} {\bibinfo
  {journal} {Phys. Rev. D}\ }\textbf {\bibinfo {volume} {100}},\ \bibinfo
  {pages} {084046} (\bibinfo {year} {2019}{\natexlab{b}})},\ \bibinfo {note}
  {[Erratum: Phys.Rev.D 101, 069902 (2020)]},\ \Eprint
  {http://arxiv.org/abs/1907.01561} {arXiv:1907.01561 [gr-qc]} \BibitemShut
  {NoStop}%
\bibitem [{\citenamefont {Maggio}\ \emph {et~al.}(2021)\citenamefont {Maggio},
  \citenamefont {van~de Meent},\ and\ \citenamefont {Pani}}]{Maggio:2021}%
  \BibitemOpen
  \bibfield  {author} {\bibinfo {author} {\bibfnamefont {E.}~\bibnamefont
  {Maggio}}, \bibinfo {author} {\bibfnamefont {M.}~\bibnamefont {van~de
  Meent}}, \ and\ \bibinfo {author} {\bibfnamefont {P.}~\bibnamefont {Pani}},\
  }\href {\doibase 10.1103/PhysRevD.104.104026} {\bibfield  {journal} {\bibinfo
   {journal} {Phys. Rev. D}\ }\textbf {\bibinfo {volume} {104}},\ \bibinfo
  {pages} {104026} (\bibinfo {year} {2021})},\ \Eprint
  {http://arxiv.org/abs/2106.07195} {arXiv:2106.07195 [gr-qc]} \BibitemShut
  {NoStop}%
\bibitem [{\citenamefont {Sago}\ and\ \citenamefont
  {Tanaka}(2021)}]{Sago:2021iku}%
  \BibitemOpen
  \bibfield  {author} {\bibinfo {author} {\bibfnamefont {N.}~\bibnamefont
  {Sago}}\ and\ \bibinfo {author} {\bibfnamefont {T.}~\bibnamefont {Tanaka}},\
  }\href {\doibase 10.1103/PhysRevD.104.064009} {\bibfield  {journal} {\bibinfo
   {journal} {Phys. Rev. D}\ }\textbf {\bibinfo {volume} {104}},\ \bibinfo
  {pages} {064009} (\bibinfo {year} {2021})},\ \Eprint
  {http://arxiv.org/abs/2106.07123} {arXiv:2106.07123 [gr-qc]} \BibitemShut
  {NoStop}%
\bibitem [{\citenamefont {Cardoso}\ and\ \citenamefont
  {Duque}(2022)}]{Cardoso:2022fbq}%
  \BibitemOpen
  \bibfield  {author} {\bibinfo {author} {\bibfnamefont {V.}~\bibnamefont
  {Cardoso}}\ and\ \bibinfo {author} {\bibfnamefont {F.}~\bibnamefont
  {Duque}},\ }\href@noop {} {\  (\bibinfo {year} {2022})},\ \Eprint
  {http://arxiv.org/abs/2204.05315} {arXiv:2204.05315 [gr-qc]} \BibitemShut
  {NoStop}%
\bibitem [{\citenamefont {Regge}\ and\ \citenamefont
  {Wheeler}(1957)}]{Regge:1957td}%
  \BibitemOpen
  \bibfield  {author} {\bibinfo {author} {\bibfnamefont {T.}~\bibnamefont
  {Regge}}\ and\ \bibinfo {author} {\bibfnamefont {J.~A.}\ \bibnamefont
  {Wheeler}},\ }\href {\doibase 10.1103/PhysRev.108.1063} {\bibfield  {journal}
  {\bibinfo  {journal} {Phys. Rev.}\ }\textbf {\bibinfo {volume} {108}},\
  \bibinfo {pages} {1063} (\bibinfo {year} {1957})}\BibitemShut {NoStop}%
\bibitem [{\citenamefont {Zerilli}(1970{\natexlab{a}})}]{Zerilli:1970wzz}%
  \BibitemOpen
  \bibfield  {author} {\bibinfo {author} {\bibfnamefont {F.~J.}\ \bibnamefont
  {Zerilli}},\ }\href {\doibase 10.1103/PhysRevD.2.2141} {\bibfield  {journal}
  {\bibinfo  {journal} {Phys. Rev. D}\ }\textbf {\bibinfo {volume} {2}},\
  \bibinfo {pages} {2141} (\bibinfo {year} {1970}{\natexlab{a}})}\BibitemShut
  {NoStop}%
\bibitem [{\citenamefont {Zerilli}(1970{\natexlab{b}})}]{Zerilli:1970se}%
  \BibitemOpen
  \bibfield  {author} {\bibinfo {author} {\bibfnamefont {F.~J.}\ \bibnamefont
  {Zerilli}},\ }\href {\doibase 10.1103/PhysRevLett.24.737} {\bibfield
  {journal} {\bibinfo  {journal} {Phys. Rev. Lett.}\ }\textbf {\bibinfo
  {volume} {24}},\ \bibinfo {pages} {737} (\bibinfo {year}
  {1970}{\natexlab{b}})}\BibitemShut {NoStop}%
\bibitem [{\citenamefont {Cardoso}\ \emph {et~al.}(2022)\citenamefont
  {Cardoso}, \citenamefont {Destounis}, \citenamefont {Duque}, \citenamefont
  {Macedo},\ and\ \citenamefont {Maselli}}]{Cardoso:2021wlq}%
  \BibitemOpen
  \bibfield  {author} {\bibinfo {author} {\bibfnamefont {V.}~\bibnamefont
  {Cardoso}}, \bibinfo {author} {\bibfnamefont {K.}~\bibnamefont {Destounis}},
  \bibinfo {author} {\bibfnamefont {F.}~\bibnamefont {Duque}}, \bibinfo
  {author} {\bibfnamefont {R.~P.}\ \bibnamefont {Macedo}}, \ and\ \bibinfo
  {author} {\bibfnamefont {A.}~\bibnamefont {Maselli}},\ }\href {\doibase
  10.1103/PhysRevD.105.L061501} {\bibfield  {journal} {\bibinfo  {journal}
  {Phys. Rev. D}\ }\textbf {\bibinfo {volume} {105}},\ \bibinfo {pages}
  {L061501} (\bibinfo {year} {2022})},\ \Eprint
  {http://arxiv.org/abs/2109.00005} {arXiv:2109.00005 [gr-qc]} \BibitemShut
  {NoStop}%
\bibitem [{\citenamefont {Panosso~Macedo}(2020)}]{PanossoMacedo:2019npm}%
  \BibitemOpen
  \bibfield  {author} {\bibinfo {author} {\bibfnamefont {R.}~\bibnamefont
  {Panosso~Macedo}},\ }\href {\doibase 10.1088/1361-6382/ab6e3e} {\bibfield
  {journal} {\bibinfo  {journal} {Class. Quant. Grav.}\ }\textbf {\bibinfo
  {volume} {37}},\ \bibinfo {pages} {065019} (\bibinfo {year} {2020})},\
  \Eprint {http://arxiv.org/abs/1910.13452} {arXiv:1910.13452 [gr-qc]}
  \BibitemShut {NoStop}%
\bibitem [{\citenamefont {Zenginoglu}\ and\ \citenamefont
  {Khanna}(2011)}]{Zenginoglu:2011zz}%
  \BibitemOpen
  \bibfield  {author} {\bibinfo {author} {\bibfnamefont {A.}~\bibnamefont
  {Zenginoglu}}\ and\ \bibinfo {author} {\bibfnamefont {G.}~\bibnamefont
  {Khanna}},\ }\href {\doibase 10.1103/PhysRevX.1.021017} {\bibfield  {journal}
  {\bibinfo  {journal} {Phys. Rev.}\ }\textbf {\bibinfo {volume} {X1}},\
  \bibinfo {pages} {021017} (\bibinfo {year} {2011})},\ \Eprint
  {http://arxiv.org/abs/1108.1816} {arXiv:1108.1816 [gr-qc]} \BibitemShut
  {NoStop}%
%%CITATION = ARXIV:1108.1816;%%
\bibitem [{\citenamefont {Krivan}\ \emph {et~al.}(1997)\citenamefont {Krivan},
  \citenamefont {Laguna}, \citenamefont {Papadopoulos},\ and\ \citenamefont
  {Andersson}}]{Krivan:1997hc}%
  \BibitemOpen
  \bibfield  {author} {\bibinfo {author} {\bibfnamefont {W.}~\bibnamefont
  {Krivan}}, \bibinfo {author} {\bibfnamefont {P.}~\bibnamefont {Laguna}},
  \bibinfo {author} {\bibfnamefont {P.}~\bibnamefont {Papadopoulos}}, \ and\
  \bibinfo {author} {\bibfnamefont {N.}~\bibnamefont {Andersson}},\ }\href
  {\doibase 10.1103/PhysRevD.56.3395} {\bibfield  {journal} {\bibinfo
  {journal} {Phys. Rev. D}\ }\textbf {\bibinfo {volume} {56}},\ \bibinfo
  {pages} {3395} (\bibinfo {year} {1997})},\ \Eprint
  {http://arxiv.org/abs/gr-qc/9702048} {arXiv:gr-qc/9702048} \BibitemShut
  {NoStop}%
\bibitem [{\citenamefont {Pazos-Avalos}\ and\ \citenamefont
  {Lousto}(2005)}]{Pazos-Avalos:2004uyd}%
  \BibitemOpen
  \bibfield  {author} {\bibinfo {author} {\bibfnamefont {E.}~\bibnamefont
  {Pazos-Avalos}}\ and\ \bibinfo {author} {\bibfnamefont {C.~O.}\ \bibnamefont
  {Lousto}},\ }\href {\doibase 10.1103/PhysRevD.72.084022} {\bibfield
  {journal} {\bibinfo  {journal} {Phys. Rev. D}\ }\textbf {\bibinfo {volume}
  {72}},\ \bibinfo {pages} {084022} (\bibinfo {year} {2005})},\ \Eprint
  {http://arxiv.org/abs/gr-qc/0409065} {arXiv:gr-qc/0409065} \BibitemShut
  {NoStop}%
\bibitem [{\citenamefont {Cardoso}\ \emph {et~al.}(2021)\citenamefont
  {Cardoso}, \citenamefont {Duque},\ and\ \citenamefont
  {Khanna}}]{Cardoso:2021vjq}%
  \BibitemOpen
  \bibfield  {author} {\bibinfo {author} {\bibfnamefont {V.}~\bibnamefont
  {Cardoso}}, \bibinfo {author} {\bibfnamefont {F.}~\bibnamefont {Duque}}, \
  and\ \bibinfo {author} {\bibfnamefont {G.}~\bibnamefont {Khanna}},\ }\href
  {\doibase 10.1103/PhysRevD.103.L081501} {\bibfield  {journal} {\bibinfo
  {journal} {Phys. Rev. D}\ }\textbf {\bibinfo {volume} {103}},\ \bibinfo
  {pages} {L081501} (\bibinfo {year} {2021})},\ \Eprint
  {http://arxiv.org/abs/2101.01186} {arXiv:2101.01186 [gr-qc]} \BibitemShut
  {NoStop}%
\bibitem [{\citenamefont {Bhattacharjee}\ \emph {et~al.}(2018)\citenamefont
  {Bhattacharjee}, \citenamefont {Mukohyama}, \citenamefont {Wan},\ and\
  \citenamefont {Wang}}]{Bhattacharjee:2018nus}%
  \BibitemOpen
  \bibfield  {author} {\bibinfo {author} {\bibfnamefont {M.}~\bibnamefont
  {Bhattacharjee}}, \bibinfo {author} {\bibfnamefont {S.}~\bibnamefont
  {Mukohyama}}, \bibinfo {author} {\bibfnamefont {M.-B.}\ \bibnamefont {Wan}},
  \ and\ \bibinfo {author} {\bibfnamefont {A.}~\bibnamefont {Wang}},\ }\href
  {\doibase 10.1103/PhysRevD.98.064010} {\bibfield  {journal} {\bibinfo
  {journal} {Phys. Rev. D}\ }\textbf {\bibinfo {volume} {98}},\ \bibinfo
  {pages} {064010} (\bibinfo {year} {2018})},\ \Eprint
  {http://arxiv.org/abs/1806.00142} {arXiv:1806.00142 [gr-qc]} \BibitemShut
  {NoStop}%
\bibitem [{\citenamefont {Mukohyama}\ and\ \citenamefont
  {Namba}(2021)}]{Mukohyama:2020lsu}%
  \BibitemOpen
  \bibfield  {author} {\bibinfo {author} {\bibfnamefont {S.}~\bibnamefont
  {Mukohyama}}\ and\ \bibinfo {author} {\bibfnamefont {R.}~\bibnamefont
  {Namba}},\ }\href {\doibase 10.1088/1475-7516/2021/02/001} {\bibfield
  {journal} {\bibinfo  {journal} {JCAP}\ }\textbf {\bibinfo {volume} {02}},\
  \bibinfo {pages} {001} (\bibinfo {year} {2021})},\ \Eprint
  {http://arxiv.org/abs/2010.09184} {arXiv:2010.09184 [hep-th]} \BibitemShut
  {NoStop}%
\bibitem [{\citenamefont {Chandrasekhar}\ and\ \citenamefont
  {Detweiler}(1975)}]{Chandrasekhar:1975zza}%
  \BibitemOpen
  \bibfield  {author} {\bibinfo {author} {\bibfnamefont {S.}~\bibnamefont
  {Chandrasekhar}}\ and\ \bibinfo {author} {\bibfnamefont {S.~L.}\ \bibnamefont
  {Detweiler}},\ }\href {\doibase 10.1098/rspa.1975.0112} {\bibfield  {journal}
  {\bibinfo  {journal} {Proc. Roy. Soc. Lond. A}\ }\textbf {\bibinfo {volume}
  {344}},\ \bibinfo {pages} {441} (\bibinfo {year} {1975})}\BibitemShut
  {NoStop}%
\bibitem [{\citenamefont {Molina}\ \emph {et~al.}(2010)\citenamefont {Molina},
  \citenamefont {Pani}, \citenamefont {Cardoso},\ and\ \citenamefont
  {Gualtieri}}]{Molina:2010fb}%
  \BibitemOpen
  \bibfield  {author} {\bibinfo {author} {\bibfnamefont {C.}~\bibnamefont
  {Molina}}, \bibinfo {author} {\bibfnamefont {P.}~\bibnamefont {Pani}},
  \bibinfo {author} {\bibfnamefont {V.}~\bibnamefont {Cardoso}}, \ and\
  \bibinfo {author} {\bibfnamefont {L.}~\bibnamefont {Gualtieri}},\ }\href
  {\doibase 10.1103/PhysRevD.81.124021} {\bibfield  {journal} {\bibinfo
  {journal} {Phys. Rev. D}\ }\textbf {\bibinfo {volume} {81}},\ \bibinfo
  {pages} {124021} (\bibinfo {year} {2010})},\ \Eprint
  {http://arxiv.org/abs/1004.4007} {arXiv:1004.4007 [gr-qc]} \BibitemShut
  {NoStop}%
\bibitem [{RDw()}]{RDwebsites}%
  \BibitemOpen
  \href@noop {} {}\bibinfo {note} {{Webpages with Mathematica notebooks and
  numerical quasinormal mode tables: \\
  \url{https://paolopani.weebly.com/notebooks.html} \\
  \url{https://centra.tecnico.ulisboa.pt/network/grit/files/} \\
  \url{https://pages.jh.edu/eberti2/ringdown/} }}\BibitemShut {NoStop}%
\bibitem [{\citenamefont {Cardoso}\ \emph
  {et~al.}(2016{\natexlab{b}})\citenamefont {Cardoso}, \citenamefont {Hopper},
  \citenamefont {Macedo}, \citenamefont {Palenzuela},\ and\ \citenamefont
  {Pani}}]{Cardoso:2016oxy}%
  \BibitemOpen
  \bibfield  {author} {\bibinfo {author} {\bibfnamefont {V.}~\bibnamefont
  {Cardoso}}, \bibinfo {author} {\bibfnamefont {S.}~\bibnamefont {Hopper}},
  \bibinfo {author} {\bibfnamefont {C.~F.~B.}\ \bibnamefont {Macedo}}, \bibinfo
  {author} {\bibfnamefont {C.}~\bibnamefont {Palenzuela}}, \ and\ \bibinfo
  {author} {\bibfnamefont {P.}~\bibnamefont {Pani}},\ }\href {\doibase
  10.1103/PhysRevD.94.084031} {\bibfield  {journal} {\bibinfo  {journal} {Phys.
  Rev. D}\ }\textbf {\bibinfo {volume} {94}},\ \bibinfo {pages} {084031}
  (\bibinfo {year} {2016}{\natexlab{b}})},\ \Eprint
  {http://arxiv.org/abs/1608.08637} {arXiv:1608.08637 [gr-qc]} \BibitemShut
  {NoStop}%
\bibitem [{\citenamefont {Berti}\ \emph {et~al.}(2016)\citenamefont {Berti},
  \citenamefont {Sesana}, \citenamefont {Barausse}, \citenamefont {Cardoso},\
  and\ \citenamefont {Belczynski}}]{Berti:2016lat}%
  \BibitemOpen
  \bibfield  {author} {\bibinfo {author} {\bibfnamefont {E.}~\bibnamefont
  {Berti}}, \bibinfo {author} {\bibfnamefont {A.}~\bibnamefont {Sesana}},
  \bibinfo {author} {\bibfnamefont {E.}~\bibnamefont {Barausse}}, \bibinfo
  {author} {\bibfnamefont {V.}~\bibnamefont {Cardoso}}, \ and\ \bibinfo
  {author} {\bibfnamefont {K.}~\bibnamefont {Belczynski}},\ }\href {\doibase
  10.1103/PhysRevLett.117.101102} {\bibfield  {journal} {\bibinfo  {journal}
  {Phys. Rev. Lett.}\ }\textbf {\bibinfo {volume} {117}},\ \bibinfo {pages}
  {101102} (\bibinfo {year} {2016})},\ \Eprint
  {http://arxiv.org/abs/1605.09286} {arXiv:1605.09286 [gr-qc]} \BibitemShut
  {NoStop}%
\bibitem [{\citenamefont {Isi}\ and\ \citenamefont {Farr}(2021)}]{Isi:2021iql}%
  \BibitemOpen
  \bibfield  {author} {\bibinfo {author} {\bibfnamefont {M.}~\bibnamefont
  {Isi}}\ and\ \bibinfo {author} {\bibfnamefont {W.~M.}\ \bibnamefont {Farr}},\
  }\href@noop {} {\  (\bibinfo {year} {2021})},\ \Eprint
  {http://arxiv.org/abs/2107.05609} {arXiv:2107.05609 [gr-qc]} \BibitemShut
  {NoStop}%
\bibitem [{\citenamefont {Ota}\ and\ \citenamefont
  {Chirenti}(2022)}]{Ota:2021ypb}%
  \BibitemOpen
  \bibfield  {author} {\bibinfo {author} {\bibfnamefont {I.}~\bibnamefont
  {Ota}}\ and\ \bibinfo {author} {\bibfnamefont {C.}~\bibnamefont {Chirenti}},\
  }\href {\doibase 10.1103/PhysRevD.105.044015} {\bibfield  {journal} {\bibinfo
   {journal} {Phys. Rev. D}\ }\textbf {\bibinfo {volume} {105}},\ \bibinfo
  {pages} {044015} (\bibinfo {year} {2022})},\ \Eprint
  {http://arxiv.org/abs/2108.01774} {arXiv:2108.01774 [gr-qc]} \BibitemShut
  {NoStop}%
\bibitem [{\citenamefont {Bhagwat}\ \emph {et~al.}(2021)\citenamefont
  {Bhagwat}, \citenamefont {Pacilio}, \citenamefont {Barausse},\ and\
  \citenamefont {Pani}}]{Bhagwat:2021kwv}%
  \BibitemOpen
  \bibfield  {author} {\bibinfo {author} {\bibfnamefont {S.}~\bibnamefont
  {Bhagwat}}, \bibinfo {author} {\bibfnamefont {C.}~\bibnamefont {Pacilio}},
  \bibinfo {author} {\bibfnamefont {E.}~\bibnamefont {Barausse}}, \ and\
  \bibinfo {author} {\bibfnamefont {P.}~\bibnamefont {Pani}},\ }\href@noop {}
  {\  (\bibinfo {year} {2021})},\ \Eprint {http://arxiv.org/abs/2201.00023}
  {arXiv:2201.00023 [gr-qc]} \BibitemShut {NoStop}%
\bibitem [{\citenamefont {Cotesta}\ \emph {et~al.}(2022)\citenamefont
  {Cotesta}, \citenamefont {Carullo}, \citenamefont {Berti},\ and\
  \citenamefont {Cardoso}}]{Cotesta:2022pci}%
  \BibitemOpen
  \bibfield  {author} {\bibinfo {author} {\bibfnamefont {R.}~\bibnamefont
  {Cotesta}}, \bibinfo {author} {\bibfnamefont {G.}~\bibnamefont {Carullo}},
  \bibinfo {author} {\bibfnamefont {E.}~\bibnamefont {Berti}}, \ and\ \bibinfo
  {author} {\bibfnamefont {V.}~\bibnamefont {Cardoso}},\ }\href@noop {} {\
  (\bibinfo {year} {2022})},\ \Eprint {http://arxiv.org/abs/2201.00822}
  {arXiv:2201.00822 [gr-qc]} \BibitemShut {NoStop}%
\end{thebibliography}%

\end{document}